\documentclass[journal]{IEEEtran}
\usepackage{mathrsfs}
\usepackage{amsfonts}
\usepackage{ifpdf}
\usepackage{cite}
\ifCLASSINFOpdf
\usepackage[pdftex]{graphicx}
\else \fi
\usepackage[cmex10]{amsmath}
\usepackage{array}
\usepackage{mdwmath}
\usepackage{mdwtab}
\usepackage{eqparbox}
\usepackage[tight,footnotesize]{subfigure}
\usepackage{algorithmic}
\usepackage{algorithm}
\usepackage{amsmath}
\usepackage{epsfig}
\usepackage{ae}
\usepackage{amssymb}
\usepackage{times}
\usepackage{amsmath}
\usepackage{amssymb}
\usepackage{graphicx}
\usepackage{booktabs}
\usepackage{color}

\usepackage{cite}

\ifCLASSINFOpdf
\else
\fi

\begin{document}

\title{Improving Physical Layer Security for Reconfigurable Intelligent Surface aided \\
NOMA 6G Networks}

\author{Zhe Zhang,%
        ~Chensi~Zhang,~\IEEEmembership{Member,~IEEE},
        ~Chengjun Jiang,
        ~Fan Jia,
        ~Jianhua Ge,
        ~and~Fengkui~Gong,~\IEEEmembership{Member,~IEEE}

\thanks{This work was supported in part by the key R \& D plan of Shaanxi Province (2019ZDLGY07-02), in part by the joint Fund of Ministry of Education of China (6141A02022338), the Fundamental Research Funds for the Central Universities, in part by the National Natural Science Foundation of China (61501347), in part by the Project Funded by China Postdoctoral Science Foundation (2015M580816), in part by the Postdoctoral Fund of Shaanxi province,in part by the ``111'' project (B08038).}%
\thanks{Z. Zhang, C. Zhang, C. Jiang, F. Jia, J. Ge and F. Gong are with the State Key Lab. of Integrated Service Networks, Xidian University, Xi'an, China. (e-mail: zhangzhe32@stu.xidian.edu.cn, cszhang@xidian.edu.cn, \{cjjiang, fjia\}@stu.xidian.edu.cn, \{jhge, fkgong\}@xidian.edu.cn).}

}

% make the title area
\maketitle

\begin{abstract}
The intrinsic integration of the nonorthogonal multiple access (NOMA) and reconfigurable intelligent surface (RIS) techniques is envisioned to be a promising approach to significantly improve both the spectrum efficiency and energy efficiency for future wireless communication networks. In this paper, the physical layer security (PLS) for a RIS-aided NOMA 6G networks is investigated, in which a RIS is deployed to assist the two ``dead zone'' NOMA users and both internal and external eavesdropping are considered. For the scenario with only internal eavesdropping, we consider the worst case that the near-end user is untrusted and may try to intercept the information of far-end user. A joint beamforming and power allocation sub-optimal scheme is proposed to improve the system PLS. Then we extend our work to a scenario with both internal and external eavesdropping. Two sub-scenarios are considered in this scenario: one is the sub-scenario without channel state information (CSI) of eavesdroppers, and another is the sub-scenario where the eavesdroppers' CSI are available. For the both sub-scenarios, a noise beamforming scheme is introduced to be against the external eavesdroppers. An optimal power allocation scheme is proposed to further improve the system physical security for the second sub-scenario. Simulation results show the superior performance of the proposed schemes. Moreover, it has also been shown that increasing the number of reflecting elements can bring more gain in secrecy performance than that of the transmit antennas.
\end{abstract}
\providecommand{\keywords}[1]{\textbf{\textit{Index terms---}} #1}

\IEEEpeerreviewmaketitle

\section{Introduction}
% no \IEEEPARstart
\IEEEPARstart{W}{ith} the rapid development of the society, industry and economy, the demand for data and access are exploding. It is forecasted that the connections will grow to 28.5 billion and the global mobile data traffic will reach 2.5 exabytes per day by 2022 \cite{r1}. From 1G to 5G, every generation of mobile communications was driven by the exponential growth of the data and access demands. Currently, in most communities, 5G networks have begun to be gradually deployed and various 5G mobile devices have appeared in the market. As one of the key technologies proposed in 5G, nonorthogonal multiple access (NOMA) have been widely investigated, owing to its ability to significantly improve the spectrum efficiency (SE) of wireless communication systems. Especially for power-domain (PD) multiplexing NOMA, multiple users are allowed to share the same spectrum resources and uses serial interference cancellation (SIC) technology to realize multi-user detection. NOMA can meet the data traffic needs of different users in the same resource block in time-domain, frequency-domain and code-domain \cite{r2}.

On the other hand, to further satisfy the traffic requirements of the emerging data-intensive applications, the relevant researchers have shown great interest in 6G technologies. The forward-looking vision of 6G had been present in \cite{r3,r4,r5}. As a new technology, reconfigurable intelligent surface (RIS)\footnote{Also known as Software Controlled Metasurface, Intelligent Wall, Passive Intelligent Mirror (PIM), Smart Reflect Array, Intelligent Reflecting Surface (IRS), Large Intelligent Surfaces (LIS) in the literature.} have attracted much attention \cite{r4,r5,r6,r7,r8}. Specifically, RIS consists of massive low-cost passive reflectors, which can reflect the signal independently by controlling its amplitude or phase to achieve passive beamforming for signal enhancement or nulling \cite{r8}. Its typical applications including wireless coverage and network throughput enhancement, interference cancelation, secure communication, wireless information and power transfer and so on. Importantly, the coverage performance can be significantly improved by properly adjusting the angle of reflection of each RIS element \cite{r9}. This is a promising approach to solve the ``dead zone'' problem in mmWave communications \cite{r8}. Particularly, compared to the conventional relays, RIS only reflects the received signals passively instead of performing the amplify-and-forward (AF) process actively, and works in full-duplex without self-interference. Thus, RIS has no additional transmission power consumption, which can effectively improve the system energy efficiency (EE) with high SE. Inspired by the merits of both NOMA and RIS, the intrinsic integration of the NOMA and RIS techniques were applied to improve the performance of the wireless system \cite{r10}.

In addition, due to the broadcast nature of the radio signals, wireless communications are more vulnerable to eavesdropping than wired ones. Therefore, physical layer security (PLS) without relying on the higher layer encryption algorithms, is getting more attention in the designs of future wireless networks \cite{r11,r12}. PLS takes advantage of the dynamic nature of wireless communication channels to ensure that legitimate users can successfully decode the data while preventing eavesdroppers from decoding the data. Compared with traditional cryptography, PLS does not require secret keys and complicated encryption processing, showing wide prospects in future wireless networks \cite{r13}.

Many relevant researchers are making outstanding contributions in aforementioned areas. Inspired by these studies, we consider a RIS-aided NOMA networks and explore the PLS performance of the system.

\subsection{Relate Work}
In recent years, RIS has received considerable attention owning to the potential and availability of this technology. The model of RIS applied to wireless networks was considered and the theoretical performance limit of RIS-aided communication system was discussed by a mathematical technique in \cite{r14}. The authors of \cite{r9} studied the free-space path loss of RIS-assisted wireless communications by developing the free-space path loss models for RIS. At first glance, RIS and relay look a little similar in application, but they are completely different concepts. The RIS-aided network and the decode-and-forward (DF) relaying aided network were compared in \cite{r15} and the results shown that the RIS has better transmission performance than DF relay when the number of RIS reflection elements is large enough. The authors of \cite{r16} presented an Energy-efficient design for RIS which makes the EE of the RIS system $300\%$ higher than that of AF relay system. In a MIMO system with RIS-aided, beamforming technique was proposed in \cite{r17,r18} to enhance the link quality of wireless communication. Robust design is important in communication systems. Under the RIS-aided communication system with imperfect CSI, the authors of \cite{r32,r33,r34} made robust transmission design under the different conditions. RIS was applied to a simultaneous wireless information and power transfer (SWIPT) aided system and multicell MIMO communications by the authors of \cite{r35,r36}, respectively.

Recently, NOMA and RIS techniques were integrated to improve both the EE and SE of the wireless networks. In \cite{r19}, RIS technique was applied in NOMA transmission to make the directions of users' channel vectors align effectively. For continuous phase shifts and discrete phase shifts of RIS elements, the authors of \cite{r20} presented different algorithms to improve the performance of the system. The downlink transmit power minimization problem for IRS-empowered NOMA network was considered in \cite{r21}. The Bit Error Rate (BER) performance of the RIS assisted NOMA system was analyzed in \cite{r22}. In \cite{r23}, RIS-assisted NOMA system was compared with traditional orthogonal multiple access (OMA) system with/without RIS and traditional NOMA system without RIS and simulation results shown that RIS-assisted NOMA system has better rate performance than others. In \cite{r37}, an RIS-assisted uplink NOMA system was considered, where the authors maximized the sum rate of all users under individual power constraint. RIS based unmanned aerial vehicles (UAV) assisted MISO NOMA downlink network was investigated in \cite{r38}.

Security has always been one of the key indicators to evaluate the quality of a communication system. To improve the PLS of a wireless network with an eavesdropper, the RIS was deployed near the eavesdropper to cancel out the effective signal received by eavesdropper in \cite{r8}, which can effectively reduce the information leakage to improve the PLS of the system. Under the constraint of secrecy rate, beamforming was used to minimize the transmitted power of the system in \cite{r24}. When the channel of eavesdropper is superior to the user's and both channels are highly correlated in space, joint beamforming was used to improve the secrecy rate of the user in \cite{r25}. The secrecy outage probability (SOP) was derived in RIS-aided wireless communication system and the effect of the number of the reflectors in the RIS on the secrecy performance was analyzed in \cite{r26}. Two algorithms were presented in \cite{r27} to enhance the PLS of RIS-aided MISO system. When the system contains multiple legitimate users and multiple eavesdroppers, a minimum-secrecy-rate maximization problem was solved to improve the secrecy performance of the whole system in \cite{r28}. In \cite{r29}, artificial noise (AN) was added into the wireless network and simulation results shown that it is beneficial to the secrecy performance of the system if AN is used appropriately. An RIS assisted MISO network with independent cooperative jamming was studied in \cite{r39} and the EE of the system was balanced while ensuring secrecy transmission.

\subsection{Motivations and Contributions}
Through the above analysis, we can observe that current research interests of RIS are the general applications of RIS, the intrinsic integration of the NOMA and RIS, and PLS of RIS-aided wireless networks. However, to our knowledge, few papers have studied the PLS for RIS-aided NOMA networks.

PD multiplexing NOMA technology enables the system to use non-orthogonal transmission at the transmitting end. Then, the receiving end detects users according to the different power of users' signals, that is, eliminate multi-access interference (MAI) through SIC. Compared with the traditional orthogonal transmission, complexity of the receiver is increased, but SE of the system can be greatly improved. In the context of 6G, RIS, as a revolutionary new technology, can improve the EE of the system by reflecting the received signals passively to the users. RIS can also solve the ``dead zone'' problem in mmWave communications. Inspired by the merits of both NOMA and RIS, we consider a NOMA system with two NOMA users which are located in the ``dead zone'' of the communication because of the obstruction between BS and users. Therefore, RIS is adopted for collaborative transmission to improve the coverage performance of the system. The RIS-aided NOMA system inherits the advantages of NOMA and RIS and is expected to be applied in 6G networks.

As is known to all, security has always been one of the key indicators during the designs of wireless communication systems. Therefore, this paper investigates a joint beamforming and power allocation scheme to improve the PLS of a RIS-aided NOMA network with two NOMA users\footnote{In the scenario with multi-users, the proposed beamforming method and power allocation scheme can also be used to adjust the order of SIC to let eavesdroppers be the preferred demodulated users.}. We first consider the scenario with internal eavesdropping and then extend our work to the scenario with both internal and external eavesdropping. For the internal eavesdropping scenario, we consider the worst case that the near-end user (NU) is untrusted and may try to intercept the information of far-end user (FU). The contributions of this paper are mainly as follows:
\begin{itemize}
  \item For the scenario with internal eavesdropping, we first establish a joint beamforming and power allocation optimization problem to improve the system PLS. A suboptimal algorithm is proposed to solve this problem. To be specific, the problem is solved in two steps, where step 1 focuses on beamforming optimization to enhance the channel of FU, and step 2 is responsible for power allocation to optimize the secrecy rate of the system. An alternate iterative algorithm is presented to obtain the beamforming vector of base station (BS) and the phase shifts of RIS in step 1. After step 1, FU is enhanced and the order of SIC could be switched, i.e., the NU was first demodulated. The simulation results show that the proposed algorithm can significantly improve the secrecy performance of the system.
  \item For the scenario with both internal and external eavesdropping, there are two sub-scenarios. One is the sub-scenario without channel state information (CSI) of eavesdroppers, another is the sub-scenario where the eavesdroppers' CSI are available. For the both sub-scenarios, a noise beamforming scheme is introduced to be against the external eavesdroppers, i.e., AN is used to prevent external eavesdroppers from eavesdropping NU. Depending on whether the eavesdroppers' CSI are available or not, two algorithms based on the Schmidt orthogonalization are presented respectively to obtain the noise beamforming matrix which can allocate AN into the null space of the channel for NOMA users. An optimal power allocation scheme is proposed to further improve the system physical security for the second sub-scenario.
  \item For the both scenarios, simulation results show that increasing the number of reflecting elements of RIS or transmit antennas of BS have a positive impact on the system secrecy performance where increasing the number of reflecting elements have larger impact than that of transmit antennas.
\end{itemize}

\subsection{Organization and Notations}
The remainder of this paper is organized as follows. In Section II, we analyze the model of RIS-aided NOMA networks without external eavesdroppers. The model of RIS-aided NOMA networks with external eavesdroppers is considered in Section III. Simulation of the model is shown in Section IV. Finally, conclusions are given in Section V.

\emph{Notations}: Superscripts ${\left(  \cdot  \right)^T}$, ${\left(  \cdot  \right)^*}$ and ${\left(  \cdot  \right)^H}$ represent transpose, conjugate and conjugate transpose, respectively. ${\rm{Diag}}\left(  \cdot  \right)$ is diagonal matrix with main diagonal $\left(  \cdot  \right)$. $\left|  \cdot  \right|$ and $\left\|  \cdot  \right\|$ represent the absolute value of scalar and 2-norm of complex vector, respectively. ${\mathbb{C}^{N \times N}}$ denotes $N \times N$ complex matrices. $x \in {\cal C}{\cal N}\left( {a,b} \right)$ represents that $x$ is complex Gaussian variable with mean $a$ and variance $b$. $y \in {\text{U}}\left( {c,d} \right)$ represents that $y$ is a $\left( {c,d} \right)$ uniformly distributed random variable. ${\rm{Rank}}\left(  \cdot  \right)$ is the rank of the matrix $\left(  \cdot  \right)$.

\section{Secrecy Design Against Internal Eavesdropping}
In this section, we consider a RIS-aided downlink NOMA transmission model with an untrusted near user. A suboptimal joint beamforming and power allocation scheme is proposed to improve the system PLS.

\subsection{System Model}
\begin{figure}[htbp]
\centering
\subfigure[]{
\includegraphics[height=0.32\textheight,angle=90]{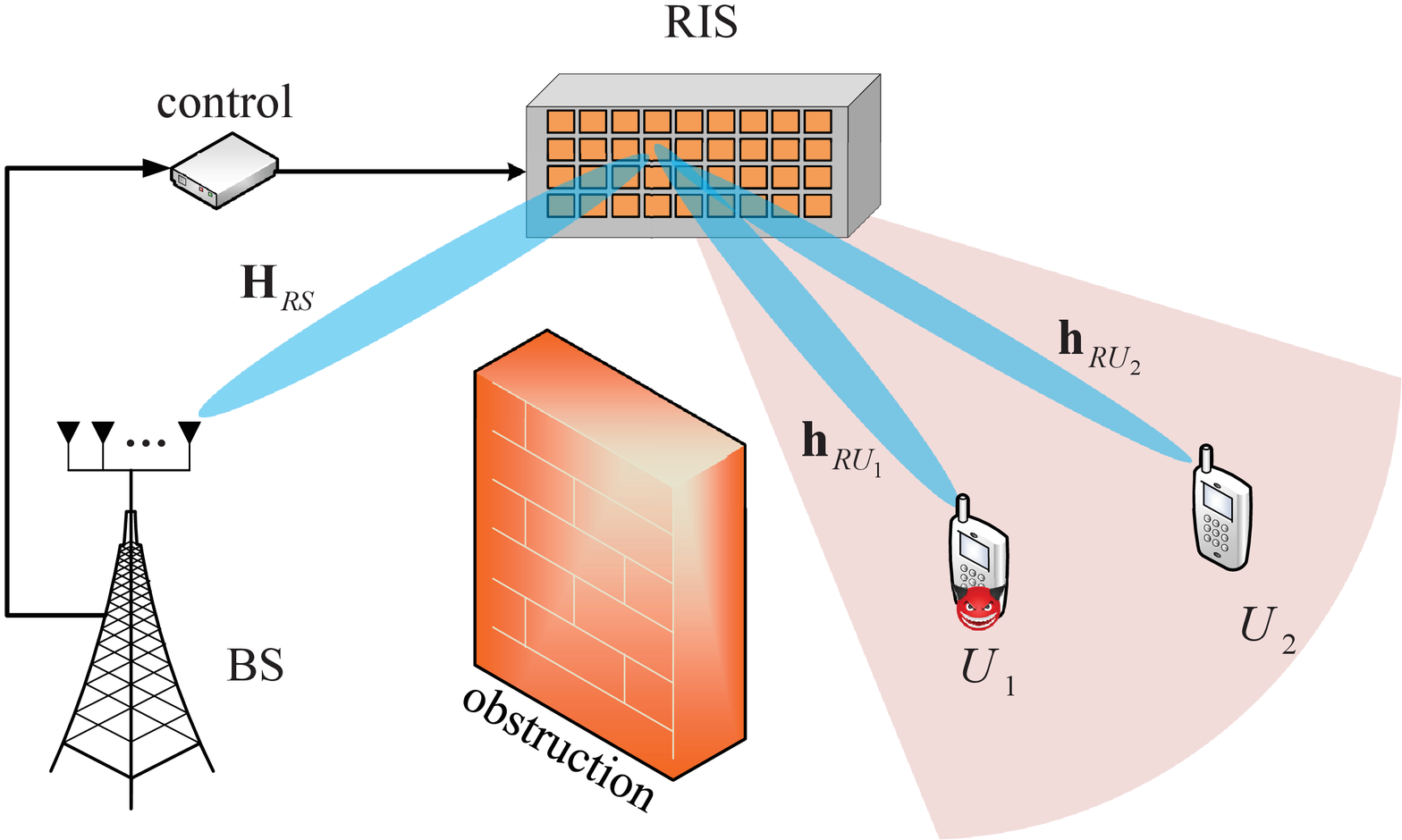}
\label{Fig_1a}
}
\quad
\subfigure[]{
\includegraphics[height=0.32\textheight,angle=90]{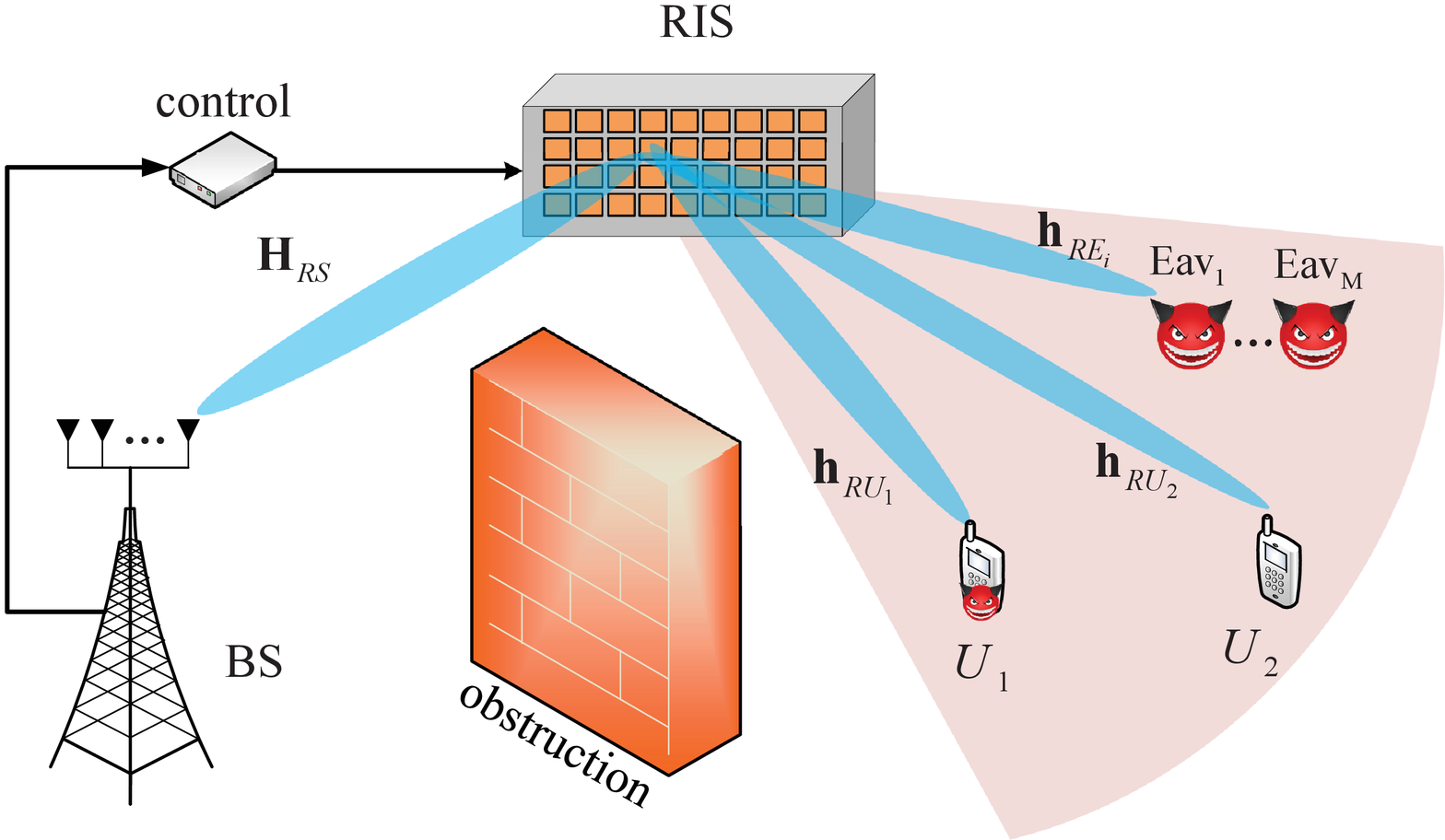}
\label{Fig_1b}
}
\caption{System model: (a) RIS-aided NOMA model with an internal eavesdropper. (b) RIS-aided NOMA model with both internal eavesdropper and external eavesdroppers.}
\end{figure}

As shown in Fig. \ref{Fig_1a}, the model consists of a BS with $Ns$ $\left( {Ns \ge 2} \right)$ transmit antennas, a RIS with $Nr$ reflecting elements and two single-antenna users ($U_1$, $U_2$), where $U_1$ is closer to RIS than $U_2$. There is no direct transmission path between BS and users because of the obstruction among themselves. The obstruction\footnote{Our system model is suitable for the signal with high frequency, such as some candidate frequency bands of mmWave include 24, 28, 39 and 60GHz. These signals with high frequency have poor diffraction performance. When the base station sends signals to the user, these signals are easily blocked by obstruction, such as the huge building in the typical urban communication system.} will block the signal from BS to users. In traditional transport schemes, relays are generally used for collaborative transmission. But relay not only consumes additional energy, but also adds additional computational complexity. In our scenario, RIS is adopted for collaborative transmission. BS sends the signal to the RIS, and the RIS reflects the signal to the users through the reflecting surface. We assume that the CSI between transport nodes can be accurately estimated. The channel coefficients of the BS-RIS link, RIS-$U_1$ link and RIS-$U_2$ link are denoted as ${{\bf{H}}_{RS}} \in {\mathbb{C}^{Nr \times Ns}}$, ${{\bf{h}}_{R{U_1}}} \in {\mathbb{C}^{Nr \times 1}}$ and ${{\bf{h}}_{R{U_2}}} \in {\mathbb{C}^{Nr \times 1}}$, respectively\footnote{The structure and element spacing of the transmit antennas have considerable impact on the channel coefficients. We assume that the elements in each channel coefficient are independently and identically distributed (i.i.d) under the design of the transmit antennas.}. The acquisition of CSI between the RIS and BS/users is discussed in \cite{r8}.

Each reflecting element of RIS will reflect the signal from the BS to the users. All of them can adjust the phase of the reflected signal by adjusting the angle of reflection. So the phase shifts applied by the RIS can be expressed as ${\bf{\Phi }} = {\rm{Diag}}\left( {{e^{j{\varphi _1}}},{e^{j{\varphi _2}}}, \cdots ,{e^{j{\varphi _{Nr}}}}} \right)$ \cite{r14}, where ${\varphi _n} \in \left[ {0,2\pi } \right]$ is phase shift of reflecting element for $n = 1,2 \cdots Nr$.

The mixed signal from BS to RIS is denoted by
\begin{equation}\label{eqn1}
{\bf{x}} = {\bf{w}}\left( {\sqrt {\left( {1 - \alpha } \right)P} {x_1} + \sqrt {\alpha P} {x_2}} \right),
\end{equation}
where $x_1$ and $x_2$ are the signal for $U_1$ and $U_2$, respectively, and {${\mathbf{w}} \in {\mathbb{C}^{Ns \times 1}}$} $\left( {\left\| {\bf{w}} \right\| = 1} \right)$ is the transmit beamformer of the BS, $P$ and $\alpha  \in (0,1)$ are transmit power and power sharing factor at BS. Then, the signals received at $U_1$ and $U_2$ can be expressed as
\begin{equation}\label{eqn2}
{y_1} = {\mathbf{h}}_{R{U_1}}^H{\mathbf{\Phi }}{{\mathbf{H}}_{RS}}{\mathbf{w}}\left( {\sqrt {\left( {1 - \alpha } \right)P} {x_1} + \sqrt {\alpha P} {x_2}} \right) + {n_1},
\end{equation}
\begin{equation}\label{eqn3}
{y_2} = {\mathbf{h}}_{R{U_2}}^H{\mathbf{\Phi }}{{\mathbf{H}}_{RS}}{\mathbf{w}}\left( {\sqrt {\left( {1 - \alpha } \right)P} {x_1} + \sqrt {\alpha P} {x_2}} \right) + {n_2},
\end{equation}
where ${n_i} \in {\cal C}{\cal N}\left( {0,{N_0}} \right)$ is the additive white Gaussian noise (AWGN) at user $i$ for $i = 1,2$.

Because $U_1$ is closer to RIS than $U_2$, the transmission channel quality between $U_1$ and RIS is superior to that between $U_2$ and RIS. Between BS and RIS, two users share a transmission channel. Therefore, the integrated transmission channel of $U_1$ is better than $U_2$. In the traditional NOMA system, BS will assign more power to $U_2$ (weak user). $U_2$ directly decodes $x_2$ by treating $x_1$ as noise. $U_1$ needs to decode $x_2$ first, and then decode $x_1$ after subtracts $x_2$ by the SIC. We consider the worst case that $U_1$ will maliciously eavesdrop on the signal of $U_2$. In this scheme, the order of SIC will be switched to protect $U_2$ from eavesdropping by $U_1$. BS will allocate more power to $U_1$. In this case, $x_1$ is the first signal to be decoded by users. As a result, the SNR at $U_1$ and $U_2$ can be respectively expressed as
\begin{equation}\label{eqn28}
{\gamma _{1,{x_1}}} = \frac{{{h_1}\left( {1 - \alpha } \right)P}}{{{h_1}\alpha P + {N_0}}},\;
{\gamma _{2,{x_1}}} = \frac{{{h_2}\left( {1 - \alpha } \right)P}}{{{h_2}\alpha P + {N_0}}},
\end{equation}
\begin{equation}\label{eqn29}
{\gamma _{1,{x_2}}} = \frac{{{h_1}\alpha P}}{{{N_0}}}
,\;
{\gamma _{2,{x_2}}} = \frac{{{h_2}\alpha P}}{{{N_0}}},
\end{equation}
where ${h_i} = {\left\| {{\bf{h}}_{R{U_i}}^H{\bf{\Phi }}{{\bf{H}}_{RS}}{\bf{w}}} \right\|^2}$ is the integrated transmission channel gain of $U_i$ and ${\gamma _{i,{x_j}}}$ is the SNR of $U_i$ to decode $x_j$ with $i,j = 1,2$. The secrecy rate of the system is given by
\begin{equation}\label{eqn30}
{R_S} = \max \left[ {0,\;{{\log }_2}\left( {1 + {\gamma _{2,{x_2}}}} \right) - {{\log }_2}\left( {1 + {\gamma _{1,{x_2}}}} \right)} \right].
\end{equation}

\subsection{Joint Beamforming and Power Allocation}
In this subsection, we propose a joint beamforming and power allocation scheme to improve the system PLS. Our goal is to maximize the secrecy rate by adjusting the value of ${\bf{\Phi }}$, ${\bf{w}}$, $\alpha $. In addition, to guarantee the communication quality of the users, data rate requirements are also taken into account. The joint beamforming and power allocation problem can be given by
\begin{equation}\label{eqn4}
\begin{gathered}
  \mathop {\max }\limits_{{\mathbf{\Phi }},{\mathbf{w}},\alpha } \;{\log _2}\left( {1 + {\gamma _{2,{x_2}}}} \right) - {\log _2}\left( {1 + {\gamma _{1,{x_2}}}} \right) \hfill \\
  {\text{s}}{\text{.t}}{\text{.}}\left\{ \begin{gathered}
  {\log _2}\left( {1 + {\gamma _{1,{x_1}}}} \right) \geqslant {R_{1,th}} \hfill \\
  {\log _2}\left( {1 + {\gamma _{2,{x_2}}}} \right) \geqslant {R_{2,th}} \hfill \\
\end{gathered}  \right. \hfill \\
\end{gathered} ,
\end{equation}
where ${R_{i,th}}$ is the minimum transmission rate threshold of $U_i$ for $i = 1,2$. However, due to the complexity of the objective function, it is hard to obtain the global optimally solution. Therefore, a suboptimal algorithm is proposed to solve this problem, which can effectively improve the secrecy performance of the system.

The problem is solved in two steps, where step 1 focuses on beamforming optimization and step 2 is responsible for power allocation:
\begin{itemize}
\item [] Step 1: We adopt a beamforming method to improve the integrated channel condition of $U_2$. Specifically, an alternate iterative algorithm is presented to determine ${\bf{\Phi }}$ and ${\bf{w}}$ to maximize the $h_2$. The optimization problem can be given by:
      \begin{equation}\label{eqn5}
      \mathop {\max }\limits_{{\bf{w}},{\bf{\Phi }}} {\rm{ }}{\left\| {{\bf{h}}_{R{U_2}}^H{\bf{\Phi }}{{\bf{H}}_{RS}}{\bf{w}}} \right\|^2}.
      \end{equation}
  \item [] Step 2: Then, we design an optimization equation to calculate the optimal power sharing factor $\alpha $. The optimization problem can be given by:
      \begin{equation}\label{eqn6}
      \begin{gathered}
      \mathop {\max }\limits_\alpha  \;\left( {\log \left( {1 + {\gamma _{2,{x_2}}}} \right) - \log \left( {1 + {\gamma _{1,{x_2}}}} \right)} \right) \hfill \\
      {\text{s}}{\text{.t}}{\text{. }}\left\{ \begin{gathered}
      \frac{{{h_1}\left( {1 - \alpha } \right)P}}{{{h_1}\alpha P + {N_0}}} \geqslant {\gamma _{1,th}} \hfill \\
      \frac{{{h_2}\alpha P}}{{{N_0}}} \geqslant {\gamma _{2,th}} \hfill \\
      \end{gathered}  \right. \hfill \\
      \end{gathered} .
      \end{equation}
      where ${\gamma _{i,th}} = {2^{{R_{i,th}}}} - 1$ for $i = 1,2$.
\end{itemize}

\emph{Step 1}: Because $U_1$ is NU, the integrated channel condition of $U_1$ is better than that of $U_2$, which leads to the secrecy rate to be zero. So, we maximize $h_2$ to enhance the channel condition of $U_2$ by adjusting ${\bf{\Phi }}$ and ${\bf{w}}$. The optimization problem is given by (\ref{eqn5}). However, it is still too difficult to obtain the closed form solution to (\ref{eqn5}). To this end, an alternate iterative algorithm is presented to find the appropriate value of ${\bf{\Phi }}$ and ${\bf{w}}$. Before applying the algorithm, we set
\begin{equation*}
\begin{gathered}
{\bf{h}}_{R{U_2}}^H\mathop  = \limits^{{\rm{def}}} \left( {{a_1}{e^{j{\beta _1}}},{a_2}{e^{j{\beta _2}}}, \cdots ,{a_{Nr}}{e^{j{\beta _{Nr}}}}} \right),\\
{{\bf{H}}_{RS}}{{\bf{w}}_k}\mathop  = \limits^{{\rm{def}}} {\left( {{b_{k1}}{e^{j{\theta _{k1}}}},{b_{k2}}{e^{j{\theta _{k2}}}}, \cdots ,{b_{kNr}}{e^{j{\theta _{kNr}}}}} \right)^T},
\end{gathered} \
\end{equation*}
where ${a_i},{b_{kj}} \in \left[ {0,\infty } \right)$, ${\beta _i},{\theta _{kj}} \in \left[ {0,2\pi } \right]$ are the amplitude and phase of corresponding vectors respectively for $i,j = 1,2, \cdots ,Nr$; $k = 0,1,2...$. The proposed alternate iterative algorithm is shown in Algorithm \ref{alg1}. The detailed derivation process in Algorithm 1 and the theoretic proof of the stationary convergence of solution are given in Appendix A.

\begin{algorithm}[htbp]
\caption{Alternate Iterative Algorithm for ${\bf{w}}$ and ${\bf{\Phi }}$}
\label{alg1}
\begin{algorithmic}[1]
\STATE  \bf{Initialization:} $\varepsilon=0.0001$, ${h_2} = 0$, $k = 0$, ${\varphi _i}=-{\beta _i}$, where $i = 1,2 \cdots ,{N_r}$;
%\fontsize{14pt}{20pt}
\STATE  ${{\bf{w}}_k} = \frac{{{{\left( {{\bf{h}}_{R{U_2}}^H{\bf{\Phi }}{{\bf{H}}_{RS}}} \right)}^H}}}{{\left\| {{\bf{h}}_{R{U_2}}^H{\bf{\Phi }}{{\bf{H}}_{RS}}} \right\|}}$;

\STATE ${h_2^k} = {\left\| {{\bf{h}}_{R{U_2}}^H{\bf{\Phi }}{{\bf{H}}_{RS}}{{\bf{w}}_k}} \right\|^2}$;

\WHILE{$\left| {{h_2^k} - {h_2}} \right| \ge \varepsilon $}
\STATE ${h_2} = {h_2^k}$;
\STATE $k = k + 1$;
\STATE ${\varphi _i} =  - {\beta _i} - {\theta _{\left( {k - 1} \right)i}}$ for $i = 1,2 \cdots ,{N_r}$;
\STATE ${{\bf{w}}_k} = \frac{{{{\left( {{\bf{h}}_{R{U_2}}^H{\bf{\Phi }}{{\bf{H}}_{RS}}} \right)}^H}}}{{\left\| {{\bf{h}}_{R{U_2}}^H{\bf{\Phi }}{{\bf{H}}_{RS}}} \right\|}}$

\STATE ${h_2^k} = {\left\| {{\bf{h}}_{R{U_2}}^H{\bf{\Phi }}{{\bf{H}}_{RS}}{{\bf{w}}_k}} \right\|^2}$;
\ENDWHILE
\STATE ${\bf{w}} = {{\bf{w}}_k}$;
\end{algorithmic}
\end{algorithm}

Obviously, the objective function is non-convex. The proof is given in Appendix B. As a result, the solution obtained by Algorithm \ref{alg1} is actually locally optimal solution. However, Algorithm \ref{alg1} has high searching efficiency, simple structure and convenient use. Moreover, simulation shows that the gap between the proposed algorithm and the optimal one is no more than $0.5\% $ with a 94 percent probability and the gap is no more than $5\% $ with a 99 percent probability. Therefore, Algorithm \ref{alg1} can be safely adopted in practical systems.

Applying Algorithm \ref{alg1}, $h_2$ is significantly enhanced. However, the event ${h_1} > {h_2}$ may still happen with a certain probability when the channel condition of $U_1$ is far better than $U_2$. When ${h_1} > {h_2}$, the secrecy rate is zero and the system is outage. In the sequel, we assume that ${h_2} > {h_1}$ always holds and the outage probability will be presented in simulation section.

\emph{Step 2}: Then, we need to optimize the power sharing factor $\alpha $. An optimization equation could be constructed as (\ref{eqn6}). By simplifying (\ref{eqn6}), we have
\begin{equation}\label{eqn7}
\begin{gathered}
  \mathop {\max }\limits_\alpha  \;\frac{{{h_2}\alpha P + {N_0}}}{{{h_1}\alpha P + {N_0}}} \hfill \\
  {\text{s}}{\text{.t}}{\text{.}}\left\{ \begin{gathered}
  \alpha  \leqslant \frac{{{h_1}P - {\gamma _{1,th}}{N_0}}}{{{h_1}P\left( {{\gamma _{1,th}} + 1} \right)}} \hfill \\
  \alpha  \geqslant \frac{{{N_0}{\gamma _{2,th}}}}{{{h_2}P}} \hfill \\
\end{gathered}  \right. \hfill \\
\end{gathered} .
\end{equation}
Observing the two constraints, it can be found that if $\frac{{{h_1}P - {\gamma _{1,th}}{N_0}}}{{{h_1}P({\gamma _{1,th}} + 1)}} < \frac{{{N_0}{\gamma _{2,th}}}}{{{h_2}P}}$, $\alpha $ has no feasible region, i.e., the problem (\ref{eqn7}) has no solution. Only when $\frac{{{h_1}P - {\gamma _{1,th}}{N_0}}}{{{h_1}P({\gamma _{1,th}} + 1)}} \ge \frac{{{N_0}{\gamma _{2,th}}}}{{{h_2}P}}$ holds, i.e.,
\begin{equation*}
P \ge ({\gamma _{1,th}} + 1)\frac{{{N_0}{\gamma _{2,th}}}}{{{h_2}}}{\rm{ + }}\frac{{{\gamma _{1,th}}}}{{{h_1}}}{N_0},
\end{equation*}
the problem has an optimal solution.

By taking the derivative of the objective function of (\ref{eqn7}) with respect to $\alpha $, it can be easily proved that the objective function is an increasing function of $\alpha $. So we can get the optimal solution, which is
\begin{equation}\label{eqn8}
{\alpha ^\Delta} = \frac{{{h_1}P - {\gamma _{1,th}}{N_0}}}{{{h_1}P({\gamma _{1,th}} + 1)}}.
\end{equation}

\section{Secrecy Design Against both Internal and External Eavesdropping}
In this section, we consider a RIS-aided downlink NOMA transmission model with both untrusted near user and $M$ unknown external eavesdroppers. Two scenarios are considered: a) the scenario without CSI of eavesdroppers; b) the scenario where the eavesdroppers' CSI are available. For the both scenarios, a noise beamforming scheme is introduced to be against the external eavesdroppers. Moreover, an optimal power allocation scheme is proposed to further improve the system physical security for the second scenario.

\subsection{System Model}
As shown in Fig. \ref{Fig_1b}, there exist both internal untrusted near user and $M$ external eavesdroppers. There is also no direct transmission path between the external eavesdroppers and the BS. The external eavesdroppers will eavesdrop on the signal reflected from RIS. We assume that external eavesdroppers are eavesdropping independently and are only interested in the $x_2$. In MIMO wireless communication system with RIS aided, AN was used to enhance the system security performance \cite{r30}. In order to prevent the external eavesdroppers from eavesdropping, we also use AN to interfere with external eavesdroppers.

${\mathbf{v}} = {\left( {{v_1},{v_2}, \cdots ,{v_{Nv}}} \right)^T}$ is the AN vector where $Nv$ is the number of the noise ${v_i} \in \mathcal{C}\mathcal{N}\left( {0,\frac{{(1 - \psi )P}}{{Nv}}} \right)$ and all the ${v_i}$ are independent of each other for $i = 1,2, \cdots ,Nv$.  The noise beamforming matrix is denoted as ${\mathbf{T}} = \left( {{{\mathbf{t}}_1},{{\mathbf{t}}_2} \cdots {{\mathbf{t}}_{Nv}}} \right)$, where ${{\mathbf{t}}_i} \in {\mathbb{C}^{Ns \times 1}}$ is unit column vector for $i = 1,2, \cdots ,Nv$.

The mixed signal sent from BS to RIS is denoted by
\begin{equation}\label{eqn9}
{\bf{x}} = {\bf{w}}\left( {\sqrt {\left( {1 - \alpha } \right)\psi P} {x_1} + \sqrt {\alpha \psi P} {x_2}} \right) + {\bf{Tv}}.
\end{equation}
$\psi  \in \left( {0,1} \right]$ is the power sharing factor between the effective signal and AN. RIS reflects $x$ from BS to the users. The signals received at $U_1$ and $U_2$ can be expressed as
\begin{equation}\label{eqn10}
\begin{split}
  {y_1} =& {\mathbf{h}}_{R{U_1}}^H{\mathbf{\Phi }}{{\mathbf{H}}_{RS}}{\mathbf{w}}\left( {\sqrt {\left( {1 - \alpha } \right)\psi P} {x_1} + \sqrt {\alpha \psi P} {x_2}} \right) \hfill \\
   +& {\mathbf{h}}_{R{U_1}}^H{\mathbf{\Phi }}{{\mathbf{H}}_{RS}}{\mathbf{Tv}} + {n_1} \hfill \\
\end{split} ,
\end{equation}
\begin{equation}\label{eqn11}
\begin{split}
  {y_2} =& {\mathbf{h}}_{R{U_2}}^H{\mathbf{\Phi }}{{\mathbf{H}}_{RS}}{\mathbf{w}}\left( {\sqrt {\left( {1 - \alpha } \right)\psi P} {x_1} + \sqrt {\alpha \psi P} {x_2}} \right) \hfill \\
   +& {\mathbf{h}}_{R{U_2}}^H{\mathbf{\Phi }}{{\mathbf{H}}_{RS}}{\mathbf{Tv}} + {n_2} \hfill \\
\end{split} ,
\end{equation}
where ${\mathbf{h}}_{R{U_i}}^H{\mathbf{\Phi }}{{\mathbf{H}}_{RS}}{\mathbf{Tv}}$ is the AN interference received by $U_i$, and other symbols have the same meaning as the model of Section II. In this model, $U_1$ will still maliciously eavesdrop on the signal of $U_2$.

\subsection{Joint Beamforming and Power Allocation without Eavesdroppers' CSI}
In this scenario, the CSI of the external eavesdroppers are not available. In this subsection, the integrated channel condition of $U_2$ is still need to be enhanced to make sure that the secrecy rate not be zero. So, ${\mathbf{\Phi }}$ and ${\mathbf{w}}$ can be achieved by Algorithm 1 in Section II. In this subsection, our main goal is to determine the noise beamforming matrix ${\mathbf{T}}$. To this end, a Schmidt orthogonalization based method is developed.

The noise beamforming matrix ${\mathbf{T}}$ is used to allocate AN into the null space of the channel for NOMA users, i.e., ${\mathbf{h}}_{R{U_i}}^H{\mathbf{\Phi }}{{\mathbf{H}}_{RS}}{\mathbf{T}} = 0$. The problem can be converted into getting $Nv$ linearly independent unit vectors ${{\mathbf{t}}_j}$ to make ${\mathbf{h}}_{R{U_i}}^H{\mathbf{\Phi }}{{\mathbf{H}}_{RS}}{{\mathbf{t}}_j} = 0$ for $j = 1,2, \cdots ,Nv$; $i = 1,2$. The author of \cite{r31} design a method to project AN into the null space of the target vector. Based on the Schmidt orthogonalization, a method denoted as Algorithm 2 is presented to find ${\mathbf{T}}$.

{\textbf {Algorithm 2}}: Set $Nv = Ns - 2$. Randomly generate $Nv$ column vectors ${{\mathbf{p}}_i}$ to make ${\rm{Rank}}\left( {{{\mathbf{c}}_1},{{\mathbf{c}}_2},{{\mathbf{p}}_1},{{\mathbf{p}}_2}, \cdots ,{{\mathbf{p}}_{Nv}}} \right) = Ns$ be true where ${{\mathbf{c}}_j} = {\left( {{\mathbf{h}}_{R{U_j}}^H{\mathbf{\Phi }}{{\mathbf{H}}_{RS}}} \right)^T}$. Then, $Nv$ unit vectors ${{\mathbf{t}}_i}$ orthogonal to ${{\mathbf{c}}_i}$ will be obtained by Schmidt orthogonalization for $i = 1,2, \cdots ,Nv$; $j = 1,2$.

Then, the signals received at $U_1$ and $U_2$ can be rewritten as
\begin{equation}\label{eqn12}
{y_1} = {\bf{h}}_{R{U_1}}^H{\bf{\Phi }}{{\bf{H}}_{RS}}{\bf{w}}\left( {\sqrt {\left( {1 - \alpha } \right)\psi P} {x_1} + \sqrt {\alpha \psi P} {x_2}} \right) + {n_1},
\end{equation}
\begin{equation}\label{eqn13}
{y_2} = {\bf{h}}_{R{U_2}}^H{\bf{\Phi }}{{\bf{H}}_{RS}}{\bf{w}}\left( {\sqrt {\left( {1 - \alpha } \right)\psi P} {x_1} + \sqrt {\alpha \psi P} {x_2}} \right) + {n_2},
\end{equation}
The SNR at $U_1$ and $U_2$ can be respectively expressed as
\begin{equation}\label{eqn31}
{\gamma _{1,{x_1}}} = \frac{{{h_1}\left( {1 - \alpha } \right)\psi P}}{{{h_1}\alpha \psi P + {N_0}}},\;
{\gamma _{1,{x_2}}} = \frac{{{h_1}\alpha \psi P}}{{{N_0}}},
\end{equation}
\begin{equation}\label{eqn32}
{\gamma _{2,{x_1}}} = \frac{{{h_2}\left( {1 - \alpha } \right)\psi P}}{{{h_2}\alpha \psi P + {N_0}}}
,\;
{\gamma _{2,{x_2}}} = \frac{{{h_2}\alpha \psi P}}{{{N_0}}}.
\end{equation}
where ${\gamma _{i,{x_j}}}$ is the SNR of $U_i$ to decode $x_j$ with $i,j = 1,2$.

We assume that the value of $\psi $ is prescribed since the eavesdroppers' CSI are not available, and the optimization problem can be given by:
\begin{equation}\label{eqn14}
\begin{gathered}
  \mathop {\max }\limits_\alpha  \;\left( {\log \left( {1 + {\gamma _{2,{x_2}}}} \right) - \log \left( {1 + {\gamma _{1,{x_2}}}} \right)} \right) \hfill \\
  {\text{s}}{\text{.t}}{\text{.}}\left\{ \begin{gathered}
  \frac{{{h_1}(1 - \alpha )\psi P}}{{{h_1}\alpha \psi P + {N_0}}} \geqslant {\gamma _{1,th}} \hfill \\
  \frac{{{h_2}\alpha \psi P}}{{{N_0}}} \geqslant {\gamma _{2,th}} \hfill \\
\end{gathered}  \right. \hfill \\
\end{gathered} .
\end{equation}
Similar to the solution of (\ref{eqn6}), only when
\begin{equation*}
P \ge ({\gamma _{1,th}} + 1)\frac{{{N_0}{\gamma _{2,th}}}}{{{h_2}\psi }}{\rm{ + }}\frac{{{\gamma _{1,th}}}}{{{h_1}\psi }}{N_0}
\end{equation*}
holds, the problem has an optimal solution, which is
\begin{equation}\label{eqn15}
{\alpha ^\Delta} = \frac{{{h_1}\psi P - {\gamma _{1,th}}{N_0}}}{{{h_1}\psi P\left( {{\gamma _{1,th}} + 1} \right)}}.
\end{equation}

\subsection{Joint Beamforming and Power Allocation with Eavesdroppers' CSI}
In this scenario, if the external eavesdroppers are untrusted legitimate users, not NOMA users, the CSI of the external eavesdroppers are available. The channel coefficient between RIS and the $i_{th}$ external eavesdropper is denoted as ${{\mathbf{h}}_{R{E_i}}} \in {\mathbb{C}^{Nr \times 1}}$ for $i = 1,2, \cdots ,M$. Similar to the previous subsection, ${\mathbf{\Phi }}$ and ${\mathbf{w}}$ can be obtained through Algorithm 1. Based on the Schmidt orthogonalization, Algorithm 3 is presented to find ${\mathbf{T}}$ which not only can allocate AN into the null space of the channel for NOMA users, but also can project AN on the channel of the external eavesdroppers as much as possible. Before applying the algorithm, set ${{\mathbf{c}}_i} = {\left( {{\mathbf{h}}_{R{U_i}}^H{\mathbf{\Phi }}{{\mathbf{H}}_{RS}}} \right)^T}$ and ${{\mathbf{d}}_j} = {\left( {{\mathbf{h}}_{R{E_j}}^H{\mathbf{\Phi }}{{\mathbf{H}}_{RS}}} \right)^T}$ for $j = 1,2, \cdots ,M$; $i = 1,2$.

{\textbf {Algorithm 3}}: According to the number of external eavesdroppers, the algorithm is divided into two cases.

\emph{Case I}: Set $Nv = M$ while ${\text{M}} \leqslant Ns - 2$ is true. For each ${{\mathbf{d}}_i}$ and ${{\mathbf{t}}_i}$, an optimization equation could be constructed as

\begin{equation*}
\begin{gathered}
  \mathop {\max }\limits_{{{\mathbf{t}}_i}} \;{{\mathbf{d}}_i}^T{{\mathbf{t}}_i} \hfill \\
  {\text{s}}{\text{.t}}{\text{.}}\left\{ \begin{gathered}
  {{\mathbf{c}}_1}^T{{\mathbf{t}}_i} = 0 \hfill \\
  {{\mathbf{c}}_2}^T{{\mathbf{t}}_i} = 0 \hfill \\
\end{gathered}  \right. \hfill \\
\end{gathered} ,
\end{equation*}
According to the knowledge of the Schmidt orthogonalization, a vector ${{\boldsymbol{\omega }}_i}$ which is orthogonal to ${{\mathbf{c}}_1}$ and ${{\mathbf{c}}_2}$ can be obtained by decomposing ${{\mathbf{d}}_i}$, i.e., ${{\mathbf{d}}_i} = {{\boldsymbol{\omega }}_i} + {k_{i1}}{{\mathbf{c}}_1} + {k_{i2}}{{\mathbf{c}}_2}$, where ${k_{ij}}$ is easy to be obtained for $i = 1,2, \cdots ,Nv$; $j = 1,2$. Then, we can get ${{\mathbf{t}}_i} = \frac{{{\boldsymbol{\omega }}_i^*}}{{\left| {{{\boldsymbol{\omega }}_i}} \right|}}$.

\emph{Case II}: Set $Nv = Ns - 2$ while ${\text{M}} > Ns - 2$ is true. Sort ${{\mathbf{d}}_i}$, i.e., if $i < j$, ${\left\| {{\mathbf{d}}_i^T{\mathbf{w}}} \right\|^2} \geqslant {\left\| {{\mathbf{d}}_j^T{\mathbf{w}}} \right\|^2}$ is true for $i = 1,2, \cdots ,M$; $j = 1,2, \cdots ,M$. With a small sequence number as priority, select $Nv$ column vectors ${{\mathbf{d}}_{{l_i}}}$ to make ${\rm{Rank}}\left( {{{\mathbf{c}}_1},{{\mathbf{c}}_2},{{\mathbf{d}}_{{l_1}}},{{\mathbf{d}}_{{l_2}}}, \cdots ,{{\mathbf{d}}_{{l_{Nv}}}}} \right) = {\rm{Rank}}\left( {{{\mathbf{c}}_1},{{\mathbf{c}}_2},{{\mathbf{d}}_1},{{\mathbf{d}}_2}, \cdots ,{{\mathbf{d}}_M}} \right)$ for $i = 1,2, \cdots ,Nv$ and $1 \leqslant {l_1} < {l_2} <  \cdots  < {l_{Nv}} \leqslant M$. By doing the same process as Case I to these column vectors, i.e., construct a series of optimization equations with each ${{\mathbf{d}}_{{l_i}}}$ and ${{\mathbf{t}}_i}$ for $i = 1,2, \cdots ,Nv$. ${\mathbf{T}}$ can be obtained by solving these optimization equations.

Then, our main goal is to maximize the secrecy rate by adjusting the power sharing factors $\alpha $ and $\psi $. There will be a high computational complexity if we consider multiple external eavesdroppers for power allocation at the same time\footnote{Similar to the Problem (\ref{eqn17}), a optimization problem can be constructed if we consider multiple external eavesdroppers for power allocation at the same time. The method to solve the problem is given as follows: according to the eavesdropper's SNR, $\psi $ can be divided into different segments in the feasible domain. Each segment has a corresponding eavesdropper playing a major role in eavesdropping. The optimal value of the secrecy rate in each segment can be obtained. The maximum of these optimal values is the solution of the problem.}. So, we only consider the eavesdropper with the best channel condition, which is denoted as $E$\footnote{The $E$ is the eavesdropper with the best channel condition, which means the channel coefficient between RIS and $E$ is ${{\mathbf{h}}_{R{E_k}}}$ where $k = \arg \mathop {\max }\limits_{i = 1 \cdots M} {\left\| {{\mathbf{h}}_{R{E_i}}^H{\mathbf{\Phi }}{{\mathbf{H}}_{RS}}{\mathbf{w}}} \right\|^2}$. In the power allocation, we only consider $E$ for external eavesdropping, but in the Section IV-B, the effect of all external eavesdroppers on the system is considered.}.

The signals received at $E$ can be expressed as
\begin{equation}\label{eqn16}
\begin{split}
  {y_E} =& {\mathbf{h}}_{RE}^H{\mathbf{\Phi }}{{\mathbf{H}}_{RS}}{\mathbf{w}}\left( {\sqrt {\left( {1 - \alpha } \right)\psi P} {x_1} + \sqrt {\alpha \psi P} {x_2}} \right) \hfill \\
   +& {\mathbf{h}}_{RE}^H{\mathbf{\Phi }}{{\mathbf{H}}_{RS}}{\mathbf{Tv}} + {n_E} \hfill \\
\end{split} ,
\end{equation}
where ${n_E} \in {\cal C}{\cal N}(0,{N_0})$ is the AWGN at $E$ and ${{\bf{h}}_{RE}} \in {\mathbb{C}^{Nr \times 1}}$ is the channel coefficients of the RIS-$E$ link. We consider the worst case that $x_1$ has being decoded by $E$. The SNR at $E$ can be expressed as
\begin{equation}\label{eqn33}
{\gamma _{E,{x_2}}} = \frac{{{h_{E1}}\alpha \psi P}}{{\frac{{\left( {1 - \psi } \right)P}}{{Nv}}{h_{E{\text{2}}}} + {N_0}}},
\end{equation}
where ${h_{E1}} = {\left\| {{\bf{h}}_{RE}^H{\bf{\Phi }}{{\bf{H}}_{RS}}{\bf{w}}} \right\|^2}$, ${h_{E2}} = {\left\| {{\bf{h}}_{RE}^H{\bf{\Phi }}{{\bf{H}}_{RS}}{\bf{T}}} \right\|^2}$. The secrecy rate of the system is given by
\begin{equation}\label{eqn34}
{R_S} = \max \left[ {0,\log \left( {1 + {\gamma _{2,{x_2}}}} \right) - \log \left( {1 + \max \left( {{\gamma _{1,{x_2}}},{\gamma _{E,{x_2}}}} \right)} \right)} \right].
\end{equation}

\begin{figure}[htbp]
\centering
\includegraphics[width=0.5\textwidth,angle=0]{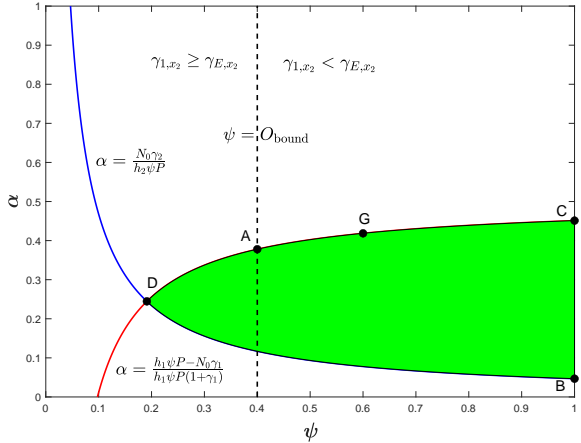}
\centering
\caption{Optimization Region for $\psi $ and $\alpha $.}
\label{Fig_2}
\end{figure}

The power sharing factors $\alpha $ and $\psi $ can be jointly optimized. The optimization problem can be given by:
\begin{equation}\label{eqn17}
\begin{gathered}
  \mathop {\max }\limits_{\psi ,\alpha } \;\left( {\log \left( {1 + {\gamma _{2,{x_2}}}} \right) - \log \left( {1 + \max \left( {{\gamma _{1,{x_2}}},{\gamma _{E,{x_2}}}} \right)} \right)} \right) \hfill \\
  {\text{s}}{\text{.t}}{\text{.}}\left\{ \begin{split}
  &\frac{{{h_1}\left( {1 - \alpha } \right)\psi P}}{{{h_1}\alpha \psi P + {N_0}}} \geqslant {\gamma _{1,th}}\;\;\;\;\;\;\;\;\;\;\;\;\;\;\;\;\;\;\;\;\;\;\;\;\;\;\;\;\;\;\;\;\;\;{\text{(}}a{\text{)}} \hfill \\
  &\frac{{{h_2}\alpha \psi P}}{{{N_0}}} \geqslant {\gamma _{2,th}}\;\;\;\;\;\;\;\;\;\;\;\;\;\;\;\;\;\;\;\;\;\;\;\;\;\;\;\;\;\;\;\;\;\;\;\;\;\;\;\;\;\;\;\;{\text{(}}b{\text{)}} \hfill \\
  &0 < \psi  \leqslant 1 \hfill \\
\end{split}  \right. \hfill \\
\end{gathered} .
\end{equation}
The boundary constructed by (\ref{eqn17}.a) and (\ref{eqn17}.b) can be expressed as
\begin{equation}\label{eqn18}
\alpha  = \frac{{{h_1}{\mkern 1mu} \psi {\mkern 1mu} P - {N_0}{\mkern 1mu} {\gamma _{1,th}}}}{{{\mkern 1mu} {h_1}{\mkern 1mu} \psi P\left( {1 + {\gamma _{1,th}}} \right)}},
\end{equation}
\begin{equation}\label{eqn19}
\alpha  = \frac{{{N_0}{\mkern 1mu} {\gamma _{2,th}}}}{{{h_2}\psi P}}.
\end{equation}
The two-dimensional optimization region for $\psi $ and $\alpha $ is shown in Fig. \ref{Fig_2}, in which the shaded region (${{\cal D}_R}$) is the feasible region of $\left( {\psi ,\alpha } \right)$. Points $B\left( {{B_x},{B_y}} \right)$, $C\left( {{C_x},{C_y}} \right)$ and $D\left( {{D_x},{D_y}} \right)$ are intersection points between the constraint boundaries. The coordinates of points in Fig. \ref{Fig_2} are shown in Appendix C.

The target function ${f_R}\left( {\psi ,\alpha } \right)$ can be expressed as:
\begin{equation}\label{eqn20}
\begin{small}
\left\{ \begin{array}{l}
\log \left( {1 + {\gamma _{2,{x_2}}}} \right) - \log \left( {1 + \max \left( {{\gamma _{1,{x_2}}},{\gamma _{E,{x_2}}}} \right)} \right),\left( {\psi {\rm{,}}\alpha } \right) \in {{\cal D}_R}\\
0,\;\;\;\;\;\;\;\;\;\;\;\;\;\;\;\;\;\;\;\;\;\;\;\;\;\;\;\;\;\;\;\;\;\;\;\;\;\;\;\;\;\;\;\;\;\;\;\;\;\;\;\;\;\;\;\;\;\;\;\;\;\;\;\;\;\;\;\;{\rm{otherwise}}
\end{array} \right..
\end{small}
\end{equation}
${C_y} \ge {B_y}$ is required to ensure $\left( {\psi ,\alpha } \right)$ has feasible region. That is, only when
\begin{equation*}
P \ge \frac{{\left( {{\gamma _{2,th}}{\mkern 1mu} \left( {1 + {\gamma _{1,th}}} \right){h_1} + {h_2}{\mkern 1mu} {\gamma _{1,th}}} \right){N_0}}}{{{h_1}{\mkern 1mu} {h_2}}}
\end{equation*}
holds, the equation has an optimal solution.

Compare the values of ${\gamma _{1,{x_2}}}$ and ${\gamma _{E,{x_2}}}$, a bound respect to $\psi $ can be found as:
\begin{equation}\label{eqn21}
{O_{{\rm{bound}}}}\mathop  = \limits^{{\rm{def}}} \frac{{\left( {{h_1} - {h_{E1}}} \right)Nv{N_0} + P{h_1}{\mkern 1mu} {h_{E2}}}}{{P{h_1}{\mkern 1mu} {h_{E2}}}}.
\end{equation}
Then, according to the value of ${O_{{\rm{bound}}}}$, the solution of problem (\ref{eqn17}) is divided into three cases
\begin{equation}\label{eqn22}
\left\{ \begin{array}{l}
{\rm{case}}\;{\rm{I}}\;\;\;\;\;\;\;\;\;\;\;\;\;\;\;\;\;\;\;\;\;\;\;1 \le {O_{{\rm{bound}}}}\\
{\rm{case}}\;{\rm{II}}\;\;\;\;\;\;\;\;\;\;\;\;\;\;\;\;\;\;\;\;\;\;{O_{{\rm{bound}}}} < {D_x}\\
{\rm{case}}\;{\rm{III}}\;\;\;\;\;\;\;\;\;\;\;\;\;\;\;\;\;\;\;\;\;{\rm{otherwise}}
\end{array} \right..
\end{equation}

\emph{Case I}: When ${O_{{\rm{bound}}}} \ge 1$, i.e., ${h_1} \ge {h_{E1}}$, ${\gamma _{1,{x_2}}} \ge {\gamma _{E,{x_2}}}$ always holds. In this case, the internal eavesdropper $U_1$ plays a major role in eavesdropping. The system model reduces to the system model without external eavesdroppers in Section II. Thus, (\ref{eqn17}) can be simplified as (\ref{eqn6}). The optimal solution is point C in the feasible region, i.e., ${\psi ^\Delta} = {C_x}$, ${\alpha ^\Delta} = {C_y}$.

\emph{Case II}: When ${O_{{\rm{bound}}}} \le {D_x}$, ${\gamma _{1,{x_2}}} \le {\gamma _{E,{x_2}}}$ is true. In this case, the external eavesdropper $E$ dominates the system PLS. Then (\ref{eqn17}) can be simplified as:
\begin{equation}\label{eqn23}
\begin{gathered}
  \mathop {\max }\limits_{\psi ,\alpha } \;\left( {1 + \frac{{{h_2}{\mkern 1mu} \alpha {\mkern 1mu} \psi {\mkern 1mu} P}}{{{N_0}}}} \right){\left( {1 + \frac{{{h_{E1}}{\mkern 1mu} \alpha {\mkern 1mu} \psi {\mkern 1mu} P}}{{\frac{{\left( {1 - \psi } \right)P{h_{E2}}}}{{Nv}} + {N_0}}}} \right)^{ - 1}} \hfill \\
  {\text{s}}{\text{.t}}{\text{.}}\left\{ \begin{gathered}
  \alpha  \leqslant \frac{{{h_1}{\mkern 1mu} \psi {\mkern 1mu} P - {N_0}{\mkern 1mu} {\gamma _{1,th}}}}{{{\mkern 1mu} {h_1}{\mkern 1mu} \psi P\left( {1 + {\gamma _{1,th}}} \right)}} \hfill \\
  \alpha  \geqslant \frac{{{N_0}{\mkern 1mu} {\gamma _{2,th}}}}{{{h_2}\psi P}} \hfill \\
  {D_x} < \psi  \leqslant 1 \hfill \\
\end{gathered}  \right. \hfill \\
\end{gathered} .
\end{equation}
By taking the partial derivative of the objective function of (\ref{eqn23}) with respect to $\alpha $, it can be easily proved that the objective function is a monotone function with respect to $\alpha $. The optimal solution of the equation must lie on the boundaries of the feasible region. There are three boundaries on the feasible region, they are (\ref{eqn18}), (\ref{eqn19}) and $\psi  = 1$, respectively.

For each boundary, the problem (\ref{eqn23}) can be solved by substituting the boundary into the objective function of  (\ref{eqn23}). Specifically, $D\left( {{D_x},{D_y}} \right)$ and $C\left( {{C_x},{C_y}} \right)$ are the optimal solution on the boundary in (\ref{eqn19}) and $\psi  = 1$, respectively. For the boundary in (\ref{eqn18}), $G\left( {{G_{\rm{x}}},{G_y}} \right)$ is a stagnation point (maximum point), and $D\left( {{D_x},{D_y}} \right)$ and $C\left( {{C_x},{C_y}} \right)$ are also boundary points.

To sum up, the optimal solution $\left( {{\psi ^\Delta},{\alpha ^\Delta}} \right)$ of the problem (\ref{eqn23}) in this case can be given by
\begin{equation}\label{eqn24}
\left\{ \begin{array}{l}
\left( {{G_{\rm{x}}},{G_y}} \right)\;\;\;\;\;\;\;\;\;\;\;\;\;\;\;\;\;\;\;\;\;\;\;\;\;\;\;\;\;\;\;\;\;\;\;\;\;\;\;\;\;\;\;\;{G_x} \in \left[ {{D_x},1} \right]\\
\arg \;\mathop {\max }\limits_{\psi ,\alpha } \left( {{f_R}\left( {{D_{\rm{x}}},{D_y}} \right),{f_R}\left( {{C_{\rm{x}}},{C_y}} \right)} \right)\;\;\;{\rm{otherwise}}
\end{array} \right..
\end{equation}

\emph{Case III}: When ${O_{{\rm{bound}}}} \in \left( {{D_x},1} \right)$, $\psi  = {O_{{\rm{bound}}}}$, a vertical dotted line in Fig. \ref{Fig_2}, divides the feasible domain into two parts: ${\gamma _{1,{x_2}}} \ge {\gamma _{E,{x_2}}}$ and ${\gamma _{1,{x_2}}} < {\gamma _{E,{x_2}}}$. Point $A\left( {{A_x},{A_y}} \right)$ is the intersection between $\psi  = {O_{{\rm{bound}}}}$ and (\ref{eqn18}). When $\psi  = {O_{{\rm{bound}}}}$ holds, ${\gamma _{1,{x_2}}} = {\gamma _{E,{x_2}}}$ is true.

In the part of ${\gamma _{1,{x_2}}} \ge {\gamma _{E,{x_2}}}$. (\ref{eqn17}) can be simplified as:
\begin{equation}\label{eqn25}
\begin{gathered}
  \mathop {\max }\limits_{\psi ,\alpha } \;\left( {\log \left( {1 + {\gamma _{2,{x_2}}}} \right) - \log \left( {1 + {\gamma _{1,{x_2}}}} \right)} \right) \hfill \\
  {\text{s}}{\text{.t}}{\text{.}}\left\{ \begin{gathered}
  \frac{{{h_1}\left( {1 - \alpha } \right)\psi P}}{{{h_1}\alpha \psi P + {N_0}}} \geqslant {\gamma _{1,th}} \hfill \\
  \frac{{{h_2}\alpha \psi P}}{{{N_0}}} \geqslant {\gamma _{2,th}} \hfill \\
  {D_x} < \psi  \leqslant {O_{{\rm{bound}}}} \hfill \\
\end{gathered}  \right. \hfill \\
\end{gathered} .
\end{equation}
It is easy to find that $A\left( {{A_x},{A_y}} \right)$ is the optimal solution in this part.

In the part of ${\gamma _{1,{x_2}}} < {\gamma _{E,{x_2}}}$, (\ref{eqn17}) can be simplified as:
\begin{equation}\label{eqn26}
\begin{gathered}
  \mathop {\max }\limits_{\psi ,\alpha } \;\left( {1 + \frac{{{h_2}{\mkern 1mu} \alpha {\mkern 1mu} \psi {\mkern 1mu} P}}{{{N_0}}}} \right){\left( {1 + \frac{{{h_{E1}}{\mkern 1mu} \alpha {\mkern 1mu} \psi {\mkern 1mu} P}}{{\frac{{\left( {1 - \psi } \right)P{h_{E2}}}}{Nv} + {N_0}}}} \right)^{ - 1}} \hfill \\
  {\text{s}}{\text{.t}}{\text{.}}\left\{ \begin{gathered}
  \alpha  \leqslant \frac{{{h_1}{\mkern 1mu} \psi {\mkern 1mu} P - {N_0}{\mkern 1mu} {\gamma _{1,th}}}}{{{\mkern 1mu} {h_1}{\mkern 1mu} \psi P\left( {1 + {\gamma _{1,th}}} \right)}} \hfill \\
  \alpha  \geqslant \frac{{{N_0}{\mkern 1mu} {\gamma _{2,th}}}}{{{h_2}\psi P}} \hfill \\
  {O_{{\text{bound}}}} < \psi  \leqslant 1 \hfill \\
\end{gathered}  \right. \hfill \\
\end{gathered} .
\end{equation}

Similar to the Case II, the optimal solution $\left( {{\psi ^\Delta},{\alpha ^\Delta}} \right)$ of the problem (\ref{eqn26}) in this case is
\begin{equation}\label{eqn27}
\left\{ \begin{array}{l}
\left( {{G_{\rm{x}}},{G_y}} \right)\;\;\;\;\;\;\;\;\;\;\;\;\;\;\;\;\;\;\;\;\;\;\;\;\;\;\;\;\;\;\;\;\;\;\;\;\;\;\;\;\;\;\;{G_x} \in \left[ {{O_{{\rm{bound}}}},1} \right]\\
\arg \;\mathop {\max }\limits_{\psi ,\alpha } \left( {{f_R}\left( {{A_{\rm{x}}},{A_y}} \right),{f_R}\left( {{C_{\rm{x}}},{C_y}} \right)} \right)\;\;{\rm{otherwise}}
\end{array} \right..
\end{equation}

\section{Simulation Results}
\begin{figure}[htbp]
\centering
\includegraphics[height=0.33\textheight,angle=90]{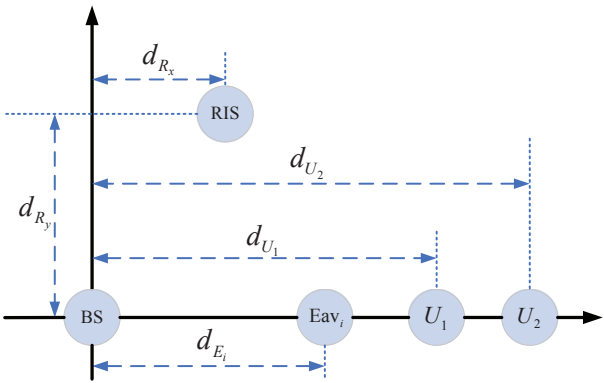}
\centering
\caption{Simulation model.}
\label{Fig_3}
\end{figure}

In this section, numerical results are presented to verify the proposed schemes. As shown in Fig. \ref{Fig_3}, cartesian coordinates is established with BS as the origin where ${\rm{Ea}}{{\rm{v}}_i}$, $U_1$ and $U_2$ are distributed in the positive X-axis and RIS is above the X-axis\footnote{For the convenience of simulation, we assume users and external eavesdroppers are stay in a line with BS. In fact, our proposed scheme can also apply to other scenarios, i.e., users and external eavesdroppers stay in arbitrary locations of plane in the manner of our model.}. ${\rm{Ea}}{{\rm{v}}_i}$ is the $i_{th}$ external eavesdropper for $i = 1,2, \cdots ,M$.

The channels from RIS to $U_1$ and $U_2$ are Los, and the users will receive not only reflected signal from RIS but also the multipath signals from the environment, so does the channel from RIS to BS. The channels between nodes are Rician channels. We assume that the $K$ factor of the Rician channel and the path loss exponent of channel are 10 and 2, respectively, e.g., ${\text{h}}_{RS}^{ij} \in \mathcal{C}\mathcal{N}\left( {\sqrt {{K \mathord{\left/
 {\vphantom {K {\left( {K + 1} \right)}}} \right.
 \kern-\nulldelimiterspace} {\left( {K + 1} \right)}}} ,{1 \mathord{\left/
 {\vphantom {1 {\left( {K + 1} \right)}}} \right.
 \kern-\nulldelimiterspace} {\left( {K + 1} \right)}}} \right)$ is the element of the ${{\mathbf{H}}_{RS}}$ of BS-RIS link for $i = 1,2, \cdots ,Nr$; $j = 1,2, \cdots ,Ns$. Then, the channel coefficients are weighted according to the path loss. We normalize the distance between nodes, i.e., set ${d_{{R_y}}} = 0.5$ and then adjust other distance parameters proportionally. We set $P = 25\;{\text{dBm}}$, ${N_0} = 0\;{\text{dBm}}$ and ${R_{1,th}} = {R_{2,th}} = 1\;{\text{bps/Hz}}$. Unless otherwise specified, these parameters are used in the following simulations. Since we care more about the relative values between the different algorithms rather than the absolute values, thus all the average secrecy rates are normalized in the following numerical results.

\subsection{Scenario with Internal Eavesdropping}
\begin{figure}[htbp]
\centering
\includegraphics[width=0.5\textwidth,angle=0]{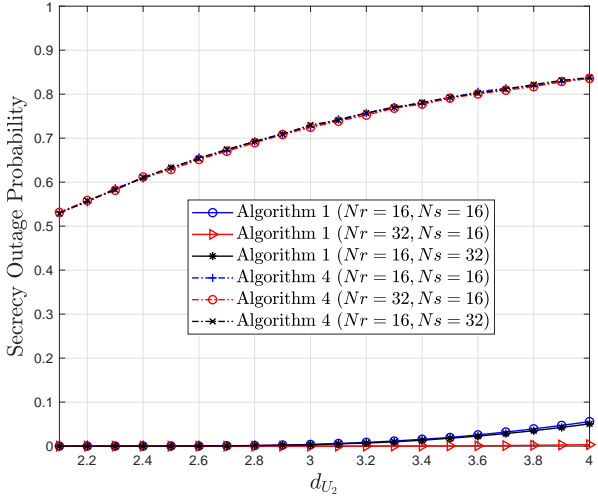}
\centering
\caption{SOP against ${d_{{U_2}}}$ with ${d_{{R_x}}} = 0.5$ and ${d_{{U_1}}} = 2$.}
\label{Fig_4}
\end{figure}

In this subsection, we consider the scenario without external eavesdropping. In this scenario, alternate iterative algorithm (see Algorithm \ref{alg1}) is proposed for beamforming (BF) to enhance the channel condition of $U_2$. In Fig. \ref{Fig_4}, we plot the SOP of the model against ${d_{{U_2}}}$ with ${d_{{R_x}}} = 0.5$ and ${d_{{U_1}}} = 2$. Additionally, the results without BF design, i.e., ${\bf{\Phi }} = {{\bf{I}}_{Nr}}$ and ${\bf{w}} = {{\bf{e}}_{Ns}}N{s^{ - \frac{{\rm{1}}}{{\rm{2}}}}}$ \footnote{${{\bf{I}}_{Nr}} \in {\mathbb{C}^{Nr \times Nr}}$ and ${{\bf{e}}_{Ns}} \in {\mathbb{C}^{Ns \times 1}}$ are the identity matrix and a column vector with all elements equal to one, respectively, i.e., ${{\bf{I}}_{Nr}} = {\rm{diag}}\left( {1,1, \cdots ,1} \right)$ and ${{\bf{e}}_{Ns}} = {\left( {1,1, \cdots ,1} \right)^T}$.}, are also presented for the comparison purpose, which is denoted as Algorithm 4. As shown in Fig. \ref{Fig_4}, the SOP increased with the increase of ${d_{{U_2}}}$ because the channel conditions for $U_2$ will deteriorate as $U_2$ moves away from RIS. The SOP obtained by Algorithm 4 is generally much higher than that of Algorithm 1, which show that the channel condition of $U_1$ is better than that of $U_2$, i.e., ${h_1} > {h_2}$ with a high probability. The result can demonstrate that the proposed algorithm could significantly enhance the channel condition of $U_2$, i.e., ${h_1} < {h_2}$ with a high probability. Moreover, from Fig. \ref{Fig_4} we can see that SOP decreases to a large extent by increasing $Nr$ from 16 to 32, while the performance gain is limited by increasing $Ns$. This observation shows that increasing the number of reflecting elements can bring more gains in secrecy performance than that of BS antennas. Finally, since Algorithm 4 does not perform beamforming operation, we can find that the values of $Nr$ and $Ns$ have little effect on the SOP of the system.

\begin{figure}[htbp]
\centering
\includegraphics[width=0.5\textwidth,angle=0]{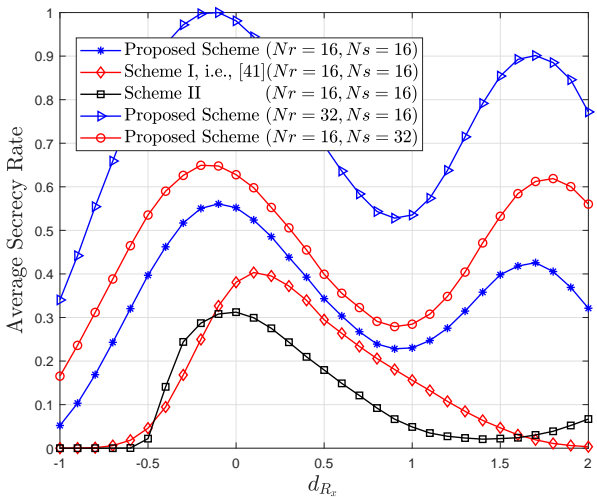}
\centering
\caption{Average secrecy rate against ${d_{{R_x}}}$ with ${d_{{U_1}}} = 1$ and ${d_{{U_2}}} = 3$.}
\label{Fig_5}
\end{figure}

The average secrecy rate under the proposed joint power allocation and beamforming scheme (Algorithm \ref{alg1}, Problem (\ref{eqn6})) are shown in Fig. \ref{Fig_5} against ${d_{{R_x}}}$ with ${d_{{U_1}}} = 1$ and ${d_{{U_2}}} = 3$. For comparison, two contrasting schemes are also considered. {For the downlink of multiuser MIMO systems, the authors of \cite{r40,r41} use the technology of space-division multiple access (SDMA) to optimize the transmission performance of the system. A transmit preprocessing technique is used to decompose the multiuser channel into multiple single-user channels without inter-user interference in \cite{r41}. According to the algorithm of \cite{r41}, transmit preprocessing vectors are used to avoid mutual eavesdropping between two users in Scheme I. In this scheme, the rest of the power is allocated to $U_2$ after meeting the minimum transmission rate threshold of $U_1$. For Scheme II, only beamforming design is adopted with fixed power allocation $\alpha  = 0.05$ (Algorithm \ref{alg1}). Fig. 5 shows that our proposed scheme is better than Scheme I. Compared with Scheme II, the result shows that our power allocation scheme can further improve the secrecy rate of the system.} Generally speaking, our scheme can effectively improve the secrecy performance compared with Scheme I and Scheme II.

The impact of the value of $Nr$ and $Ns$ on the system is also provided in Fig. \ref{Fig_5}. Similarly, we can find that more performance gain can be achieved by increasing the number of reflecting elements (i.e., $Nr$). Moreover, we can also observe that there are three extreme points in our proposed scheme, one of which is the local minimum point and the others are the local maximum points. When RIS moves to the vicinity of $U_1$ or BS, the average secrecy rate reaches local minimum or local maximum, respectively. This observation can provide a useful guideline for the deployment of RIS.

\begin{figure}[htbp]
\centering
\includegraphics[width=0.5\textwidth,angle=0]{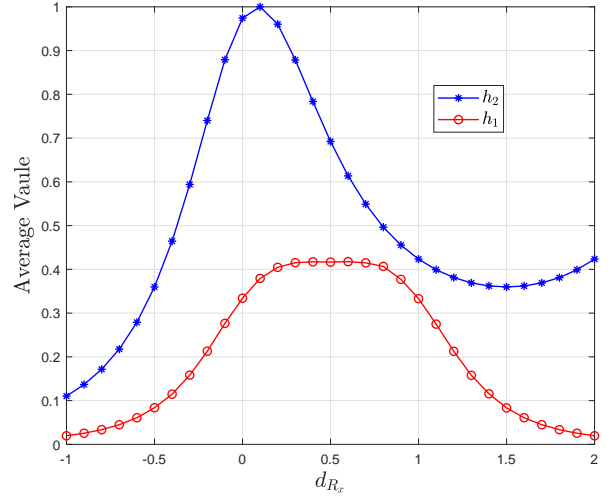}
\centering
\caption{Average value against ${d_{{R_x}}}$ with ${d_{{U_1}}} = 1$ and ${d_{{U_2}}} = 3$.}
\label{Fig_6}
\end{figure}

{In our model, the performance of the system is greatly affected by the values of ${h_1}$ and ${h_2}$, especially the difference value between them. Therefore, relative mean values of ${h_1}$ and ${h_2}$ are simulated to analyze the trend of system performance in Fig. \ref{Fig_5}. Based on the Fig. \ref{Fig_5}, the average value of ${h_1}$ and ${h_2}$ against ${d_{{R_x}}}$ are shown in Fig. \ref{Fig_6}. With the influence of Algorithm 1, the channel condition of $U_1$ is optimal when RIS is near the BS, that is, ${h_2}$ reaches the maximum. ${h_1}$ and ${h_2}$ decrease when the RIS is away from the BS to the left, resulting in a decline in system performance. When the RIS moves from BS to $U_1$, ${h_2}$  decreases while ${h_1}$ remains flat. And when the RIS moves from $U_1$ to $U_2$, the downward trend of $h_2$ gradually flattens out and even increases, while ${h_1}$ starts to decrease. But when RIS moves to the right by more than a value, the value of ${h_1}$ cannot meet the minimum transmission rate threshold of ${U_1}$ because of the limitation of power. In this case, the performance of system decreased. The result shows that our analysis is consistent with the trend of system performance in Fig. \ref{Fig_5}.}

\begin{figure}[htbp]
\centering
\includegraphics[width=0.5\textwidth,angle=0]{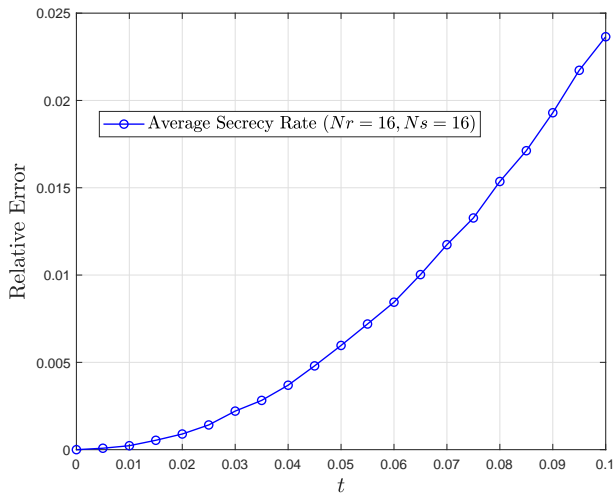}
\centering
\caption{Relative error against $t$ with ${d_{{R_x}}} = 0.5$, ${d_{{U_1}}} = 2$ and ${d_{{U_2}}} = 3$.}
\label{Fig_7}
\end{figure}

{In this article, we assume that all the CSI of channels between nodes are perfect. But it is a challenge to obtain the perfect CSI due to the time-varying characteristics of the channel and the limited signal processing capability of RIS. The authors of \cite{r32,r33,r34} studied the robust performance optimization method of RIS under the condition with imperfect CSI. Our proposed scheme is also suitable for the scenario with imperfect CSI. But the imperfect CSI still has an impact on the results obtained by our proposed scheme. It is necessary to analyze the secrecy performance of the system with imperfect CSI. Therefore, we consider a scenario with imperfect CSI. Specifically, there are errors in the CSI between transport nodes. The estimated channel can be modeled as $\mathop h\limits^ \wedge   = h + e$, where $h$ is the real CSI of channel and $e \in \mathcal{C}\mathcal{N}\left( {0,\sigma _e^2} \right)$ is the error of estimation \cite{r42}. The variance of $h$ is denoted as $\sigma _h^2$. We set $t = \sqrt {{{\sigma _e^2} \mathord{\left/
 {\vphantom {{\sigma _e^2} {\sigma _h^2}}} \right.
 \kern-\nulldelimiterspace} {\sigma _h^2}}} $ to determine the value of $e$. Base on Fig. \ref{Fig_5}, our proposed scheme is applied to the scenario with perfect CSI and imperfect CSI, respectively. The relative error against $t$ with ${d_{{R_x}}} = 0.5$, ${d_{{U_1}}} = 2$ and ${d_{{U_2}}} = 3$ are shown in Fig. \ref{Fig_7}. We can see that the relative error of the system increased with the increase of $t$. But, when the estimation error reaches $0.1$, the relative error is only $0.025$, i.e., there is small relative error in our system. The result shows that our proposed scheme is suitable for the scenario with imperfect CSI.}

\subsection{Scenario with both Internal and External Eavesdropping}
In this subsection, the scenario with both internal and external eavesdropping is considered. AN is used to interfere with external eavesdroppers. We assume that there are 10 eavesdroppers in the simulation model, i.e., $M = 10$. The eavesdropper is uniformly distributed in a certain range, i.e., ${d_{{E_i}}} \in {\text{U}}\left( {1,1.5} \right)$ for $i = 1,2, \cdots ,M$.

\begin{figure}[htbp]
\centering
\includegraphics[width=0.5\textwidth,angle=0]{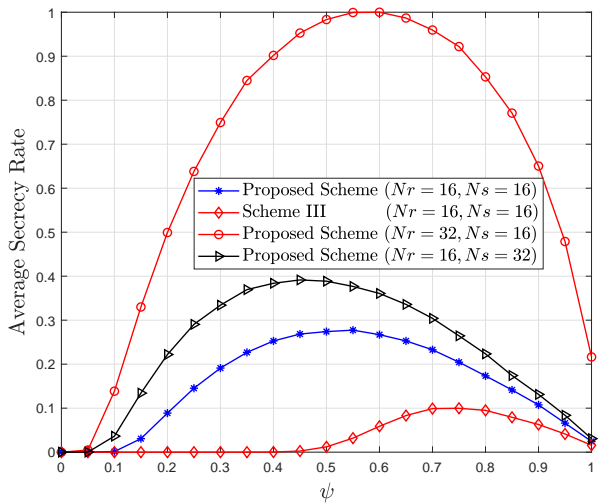}
\centering
\caption{Average secrecy rate against $\psi $ with ${d_{{R_x}}} = 0.5$ and ${d_{{U_1}}} = 2$, ${d_{{U_2}}} = 3$.}
\label{Fig_8}
\end{figure}

In the scenario without eavesdroppers' CSI, Algorithm 2 is used to generate noise beamforming matrix ${\mathbf{T}}$ to assist transmission for all the schemes in this scenario. The average secrecy rate against $\psi $ with ${d_{{R_x}}} = 0.5$ and ${d_{{U_1}}} = 2$, ${d_{{U_2}}} = 3$ are shown in the Fig. \ref{Fig_8}. In Fig. \ref{Fig_8}, the proposed scheme refers to joint power allocation and beamforming scheme (Algorithm \ref{alg1}, Problem (\ref{eqn14})), and Scheme III refers to the scheme with only BF (i.e., Algorithm \ref{alg1}) with fixed power allocation $\alpha  = 0.05$. As shown in Fig. \ref{Fig_8}, power allocation designs can significantly increase the system performance, especially for the case where RIS equipped more reflecting elements. We can see from Fig. \ref{Fig_8} that the point of $\psi  = 1$, i.e., AN is not used in this point, is lower than most other points. Therefore, it is necessary to use AN to improve the secrecy performance of the system.

\begin{figure}[htbp]
\centering
\includegraphics[width=0.5\textwidth,angle=0]{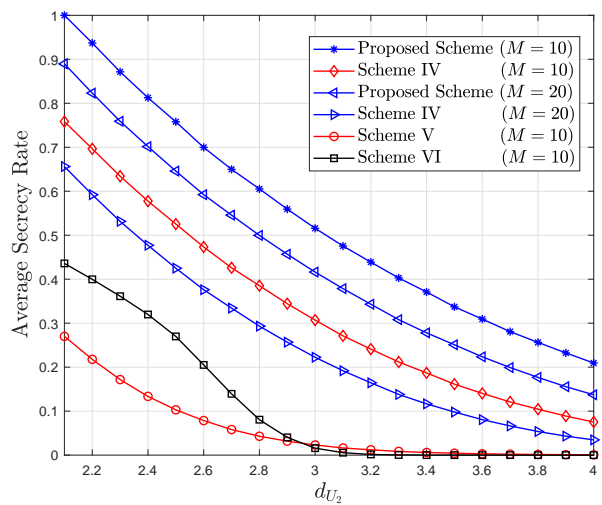}
\centering
\caption{Average secrecy rate against ${d_{{U_2}}}$ with ${d_{{R_x}}} = 0.5$, ${d_{{U_1}}} = 2$ and $Ns = Nr = 16$.}
\label{Fig_9}
\end{figure}

In addition, from Fig. \ref{Fig_8} we found that the performance undergone great changes against $\psi $, illustrating that the value of $\psi $ has a big impact on secrecy performance. Therefore, it is necessary to optimize $\psi $ when the CSI of eavesdroppers are known. In the scenario with eavesdroppers' CSI, the average secrecy rate against ${d_{{U_2}}}$ with ${d_{{R_x}}} = 0.5$, ${d_{{U_1}}} = 2$ and $Ns = Nr = 16$ are shown in Fig. \ref{Fig_9}. In Fig. \ref{Fig_9}, the proposed scheme refers to joint power allocation and beamforming scheme (Algorithm \ref{alg1}, Problem (\ref{eqn17})), and Algorithm 3 is used to generate noise beamforming matrix ${\mathbf{T}}$. Base on the proposed scheme, Algorithm 2 is used to generate ${\mathbf{T}}$ in Scheme IV to prove the better performance of Algorithm 3. AN is not used in Scheme V which refers to the scheme with power allocation and beamforming scheme in Section II (Algorithm \ref{alg1}, Problem (\ref{eqn6})). Scheme VI refers to the scheme with BF (i.e., Algorithm \ref{alg1}) and Algorithm 3 with fixed power allocation $\alpha  = 0.05$, $\psi  = 0.5$. {In Algorithm 3, according to the number of external eavesdroppers, the algorithm is divided into two cases. In order to fully show the performance of the two cases in Algorithm 3, we set $M = 10$ and $M = 20$ respectively in the simulation, i.e. $M \leqslant Ns - 2$ (Algorithm 3, case I) and $M > Ns - 2$ (Algorithm 3, case II).} We can see from Fig. \ref{Fig_9} that Algorithm 3 has better secrecy performance than Algorithm 2 by comparing with the Scheme IV and the performance of the system deteriorated with the increase of eavesdroppers. By comparing Scheme V and Scheme VI, we can find that AN and joint power allocation can improve the secrecy performance to a great extent.

\begin{figure}[htbp]
\centering
\includegraphics[width=0.5\textwidth,angle=0]{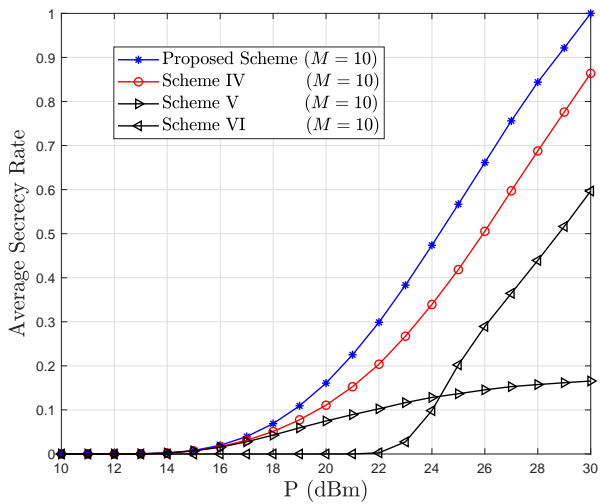}
\centering
\caption{Average secrecy rate against $P$ with ${d_{{R_x}}} = 0.5$, ${d_{{R_y}}} = 1$ and $Ns = Nr = 16$.}
\label{Fig_10}
\end{figure}

Then, we consider a scenario with dynamic users, i.e., users and eavesdroppers move randomly over a range, rather than standing still at a certain point. Instead of being confined to a line with BS, they move randomly in the plane. In this scenario, eavesdroppers, $U_1$ and $U_2$ are uniformly distributed in the circle with $\left( {{\text{2}},0} \right)$, $\left( {{\text{3}},0} \right)$ and $\left( {{\text{4}},0} \right)$ as the center and 0.5 as the radius, respectively. The average secrecy rate against $P$ with ${d_{{R_x}}} = 0.5$, ${d_{{R_y}}} = 1$ and $Ns = Nr = 16$ are shown in Fig. \ref{Fig_10}. We can see that the performance of the system increased with the increase of $P$. When the power increases to a certain value, the power gain of scheme V is obviously lower than that of other schemes. Because if we do not use AN in the system, it is no longer the power that limits the performance of the system, but the impact of external eavesdropping on the system. We can find from Fig. \ref{Fig_10} that our proposed scheme can apply to the scenario with dynamic users and can significantly improve the secrecy performance of the system.

\section{Conclusions}
In this paper, PLS for RIS-aided NOMA system was investigated. We first consider the scenario with only internal eavesdropping, i.e., NU is untrusted and may try to intercept the information of FU. A joint beamforming and power allocation scheme was proposed to improve the PLS of the system. Due to the complexity of the problem, a suboptimal joint beamforming and power allocation scheme was proposed to solve the question. Then, our work was extended to the scenario with both internal and external eavesdropping. There are two sub-scenarios: one is the sub-scenarios without CSI of eavesdroppers, another is the sub-scenarios where the eavesdroppers' CSI are available. For the both sub-scenarios, AN is used to prevent the external eavesdroppers from eavesdropping. Depending on whether the eavesdroppers' CSI are available or not, two algorithms based on the Schmidt orthogonalization are presented respectively to obtain the noise beamforming matrix which can allocate AN into the null space of the channel for NOMA users. In addition, an optimal power allocation scheme is proposed to optimally allocate the power for jamming and transmitting signals for the second sub-scenario. It has been also shown that increasing the number of reflecting elements of RIS or transmit antennas of BS will improve the secrecy performance of the system. Specifically, increasing the number of reflecting elements can bring more gain in secrecy performance than that of transmit antennas. {The simulation result shows that there is small error of our proposed scheme in the scenario with imperfect CSI, which prove that our scheme is suitable for the scenario.}

\begin{appendices}
\section{}
The detailed derivation process in Algorithm 1 is given as follows:

We regard ${\mathbf{\Phi }}$ and ${\mathbf{w}}$ as variables of two different dimensions, respectively. First, the initial value is assigned to the ${\mathbf{\Phi }}$, i.e., ${\varphi _i} =  - {\beta _i}$ for $i = 1,2 \cdots ,{N_r}$. Then, take ${\mathbf{w}}$ as the variable to obtain the maximum value of objective function ${\left\| {{\mathbf{h}}_{R{U_2}}^H{\mathbf{\Phi }}{{\mathbf{H}}_{RS}}{\mathbf{w}}} \right\|^2}$, i.e.,
\begin{equation*}
{{\mathbf{w}}_0} = \arg \mathop {\max }\limits_{\mathbf{w}} {\left\| {{\mathbf{h}}_{R{U_2}}^H{\mathbf{\Phi }}{{\mathbf{H}}_{RS}}{\mathbf{w}}} \right\|^2} = \frac{{{{\left( {{{\mathbf{h}}_{R{U_2}}}^H{\mathbf{\Phi }}{{\mathbf{H}}_{RS}}} \right)}^H}}}{{\left\| {{{\mathbf{h}}_{R{U_2}}}^H{\mathbf{\Phi }}{{\mathbf{H}}_{RS}}} \right\|}}.
\end{equation*}
Then alternate iteration loop is performed for ${\mathbf{\Phi }}$ and ${\mathbf{w}}$. In the $k_{th}$ loop, a new value of ${\mathbf{\Phi }}$ can be obtained by
\begin{equation*}
\mathop {\max }\limits_{\mathbf{\Phi }} {\left\| {{\mathbf{h}}_{R{U_2}}^H{\mathbf{\Phi }}{{\mathbf{H}}_{RS}}{{\mathbf{w}}_{k - 1}}} \right\|^2},
\end{equation*}
i.e., ${\varphi _i} =  - {\beta _i} - {\theta _{\left( {k - 1} \right)i}}$ for $i = 1,2 \cdots ,{N_r}$ and then we can get
\begin{equation*}
{{\mathbf{w}}_k} = \arg \mathop {\max }\limits_{\mathbf{w}} {\left\| {{\mathbf{h}}_{R{U_2}}^H{\mathbf{\Phi }}{{\mathbf{H}}_{RS}}{\mathbf{w}}} \right\|^2} = \frac{{{{\left( {{{\mathbf{h}}_{R{U_2}}}^H{\mathbf{\Phi }}{{\mathbf{H}}_{RS}}} \right)}^H}}}{{\left\| {{{\mathbf{h}}_{R{U_2}}}^H{\mathbf{\Phi }}{{\mathbf{H}}_{RS}}} \right\|}}.
\end{equation*}
 Execute the loop until $\left| {h_2^k - h_2^{k - 1}} \right| < \varepsilon $ is true where $h_2^k = {\left\| {{\mathbf{h}}_{R{U_2}}^H{\mathbf{\Phi }}{{\mathbf{H}}_{RS}}{{\mathbf{w}}_k}} \right\|^2}$ for $k = 1,2, \cdots $. The accuracy of error is denoted as $\varepsilon $. When the loop is completed, the resulting ${\mathbf{\Phi }}$ and ${{\mathbf{w}}_k}$ are the phase shifts matrix and beamforming vector of the system.

It is necessary to prove the convergence of the algorithm results, i.e., prove that the sequence $h_2^k$ is convergent. It is well known that if a sequence is increasing and bounded above, then it must be convergent. From the algorithm derivation, it is easy to find that the sequence $h_2^k$ is increasing. According to the nature of the norm,
\begin{equation*}
0 \leqslant {\left\| {{\mathbf{h}}_{R{U_2}}^H{\mathbf{\Phi }}{{\mathbf{H}}_{RS}}{\mathbf{w}}} \right\|^2} \leqslant {\left\| {{\mathbf{h}}_{R{U_2}}^H{\mathbf{\Phi }}} \right\|^2}{\left\| {{{\mathbf{H}}_{RS}}{\mathbf{w}}} \right\|^2}
\end{equation*}
is true. It is easy to find that ${\left\| {{\mathbf{h}}_{R{U_2}}^H{\mathbf{\Phi }}} \right\|^2}$ and ${\left\| {{{\mathbf{H}}_{RS}}{\mathbf{w}}} \right\|^2}$ are bounded in their feasible region, respectively. So ${\left\| {{\mathbf{h}}_{R{U_2}}^H{\mathbf{\Phi }}{{\mathbf{H}}_{RS}}{\mathbf{w}}} \right\|^2}$ is bounded, i.e., $h_2^k$ has an upper bound. So, the sequence $h_2^k$ is convergent.

\section{}
It is hard to directly prove that the objective function is non-convex because of massive variables. So, we treat all variables as constants except ${\varphi _i}$, where ${\varphi _i}$ is the phase shift of the $i_{th}$ element of RIS. Then, the objective function ${\left\| {{\mathbf{h}}_{R{U_2}}^H{\mathbf{\Phi }}{{\mathbf{H}}_{RS}}{\mathbf{w}}} \right\|^2}$ can be simplified as ${\left\| {{q_1}{e^{j{\varphi _i}}} + {q_2}} \right\|^2}$ where ${q_k}$ is a constant for $k = 1,2$. We can find a ${\varphi _i}$ to make ${q_1} \ne 0$ be true. Further, function can be simplified as:
\begin{equation*}
\begin{gathered}
  \;\;\;{\left\| {{q_1}{e^{j{\varphi _i}}} + {q_2}} \right\|^2} \hfill \\
   = {\left( {{q_1}{e^{j{\varphi _i}}} + {q_2}} \right)^H}\left( {{q_1}{e^{j{\varphi _i}}} + {q_2}} \right) \hfill \\
   = {\left\| {{q_1}} \right\|^2} + {\left\| {{q_2}} \right\|^2} + \operatorname{Re} \left( {{q_1}q_2^H{e^{j{\varphi _i}}}} \right) \hfill \\
   = {\left\| {{q_1}} \right\|^2} + {\left\| {{q_2}} \right\|^2} + \tau \operatorname{Re} \left( {{e^{j\left( {\mu  + {\varphi _i}} \right)}}} \right) \hfill \\
   = {\left\| {{q_1}} \right\|^2} + {\left\| {{q_2}} \right\|^2} + \tau \cos \left( {\mu  + {\varphi _i}} \right) \hfill \\
\end{gathered}
\end{equation*}
where ${q_1}q_2^H = \tau {e^{j\mu }}$. Obviously, the function is non-convex in range of ${\varphi _i} \in \left( {0,2\pi } \right)$. So, the objective function ${\left\| {{\mathbf{h}}_{R{U_2}}^H{\mathbf{\Phi }}{{\mathbf{H}}_{RS}}{\mathbf{w}}} \right\|^2}$ is non-convex.

\section{}
The coordinates of points in Fig. \ref{Fig_2} are shown as follow:

\begin{flalign*}
&{A_x} = \frac{{\left( {{h_1} - {h_{E1}}} \right)Nv{N_0} + P{h_1}{\mkern 1mu} {h_{E2}}}}{{P{h_1}{\mkern 1mu} {h_{E2}}}}.&
\end{flalign*}
\begin{flalign*}
&
\begin{gathered}
{A_y} = \left(\left( {Nv{h_1} - {Nv}{h_{E1}} - {h_{E2}}{\mkern 1mu} {\gamma _{1,th}}} \right){N_0} \right. \hfill \\
\;\;\;\;\; +\left. P{h_1}{\mkern 1mu} {h_{E2}}\right){\left( {\left( {{h_1} - {h_{E1}}} \right)Nv{N_0} + P{h_1}{\mkern 1mu} {h_{E2}}} \right)^{ - 1}} \hfill \\
\;\;\;\;\;{\times \left( {1 + {\gamma _{1,th}}} \right)^{ - 1}} .\hfill
\end{gathered}
&
\end{flalign*}
\begin{flalign*}
&B\left( {{B_x},{B_y}} \right) = \left( {1,\frac{{{N_0}{\mkern 1mu} {\gamma _{2,th}}}}{{{h_2}{\mkern 1mu} P}}} \right). \hfill&
\end{flalign*}
\begin{flalign*}
&C\left( {{C_x},{C_y}} \right) = \left( {1,\frac{{P{h_1} - {N_0}{\mkern 1mu} {\gamma _{1,th}}}}{{{h_1}{\mkern 1mu} P\left( {1 + {\gamma _{1,th}}} \right)}}} \right). \hfill&
\end{flalign*}
\begin{flalign*}
&{D_x} = \frac{{\left( {{\gamma _{2,th}}{\mkern 1mu} \left( {1 + {\gamma _{1,th}}} \right){h_1} + {h_2}{\mkern 1mu} {\gamma _{1,th}}} \right){N_0}}}{{{h_1}{\mkern 1mu} {h_2}{\mkern 1mu} P}}. \hfill&
\end{flalign*}
\begin{flalign*}
&{D_y} = \frac{{{\gamma _{2,th}}{\mkern 1mu} {h_1}}}{{{\gamma _{2,th}}{\mkern 1mu} \left( {1 + {\gamma _{1,th}}} \right){h_1} + {h_2}{\mkern 1mu} {\gamma _{1,th}}}} .\hfill&
\end{flalign*}
\begin{flalign*}
&
\begin{gathered}
    {G_x} = \left( - P{\mkern 1mu} {h_1}{\mkern 1mu} {h_2}{\mkern 1mu} \left( {1 + {\gamma _{1,th}}} \right)h_{E2}^2 - {h_2}{\mkern 1mu} \left( {\left( {1 + {\gamma _{1,th}}} \right){h_1}} \right. \right. \hfill \\
    \;\;\;\;\;\left. - {h_{E1}}{\mkern 1mu} {\gamma _{1,th}}\right) Nv{N_0}{\mkern 1mu} {h_{E2}} + \left({h_2}{\mkern 1mu} {h_{E1}}{\mkern 1mu} {h_{E2}}{\mkern 1mu} \left( {1 } \right. \right. \hfill \\
    \;\;\;\;\; \left.+ {\gamma _{1,th}}\right) Nv\left( {\left( { - {N_0}{\mkern 1mu} {\gamma _{1,th}} + {h_1}{\mkern 1mu} P} \right){h_{E2}} + {N_0}{\mkern 1mu} {h_1}{\mkern 1mu} } \right. \hfill \\
    \;\;\;\;\;\left.\times Nv\right) \left({h_{E2}}\left( {\left( { - {N_0}{\mkern 1mu} {\gamma _{1,th}} + {h_1}{\mkern 1mu} P} \right){h_2} + {N_0}{\mkern 1mu} {h_1}{\mkern 1mu} } \right. \right. \hfill \\
    \;\;\;\;\left. \left. \left.  \times\left. \left( {1 + {\gamma _{1,th}}} \right)\right)+{N_0}{\mkern 1mu} {h_1}\left( {{h_2} - {h_{E1}}} \right)Nv\right) \right)^{\frac{1}{2}}\right) \hfill \\
    \;\;\;\;\;\times {\left( {{h_2}{\mkern 1mu} P{h_{E2}}{\mkern 1mu} {h_1}{\mkern 1mu} \left( {\left( { - {\gamma _{1,th}} - 1} \right){h_{E2}} + {h_{E1}}{\mkern 1mu} Nv} \right)} \right)^{ - 1}}.\hfill
\end{gathered}
&
\end{flalign*}
\begin{flalign*}
&
\begin{gathered}
    {G_y} = \left( - {h_2}{\mkern 1mu} \left( {1 + {\gamma _{1,th}}} \right)\left( { - {N_0}{\mkern 1mu} {\gamma _{1,th}} + {h_1}{\mkern 1mu} P} \right)h_{E2}^2 - {N_0}{\mkern 1mu} {h_1}{\mkern 1mu} {h_2}{\mkern 1mu}  \right. \hfill \\
    \;\;\;\;\;\;\times {h_{E2}}\left( {1 + {\gamma _{1,th}}} \right)Nv + \left(\left( {1 + {\gamma _{1,th}}} \right){h_{E1}}{\mkern 1mu} {h_2}{\mkern 1mu} {h_{E2}}{\mkern 1mu}  \right. \hfill \\
    \;\;\;\;\;\;\times Nv\left( {\left( {Nv{N_0} + P{h_{E2}}} \right){h_1} - {N_0}{\mkern 1mu} {h_{E2}}{\mkern 1mu} {\gamma _{1,th}}} \right) \hfill \\
    \;\;\;\;\;\;\times \left({h_1}\left(\left( {{h_{E2}}{\mkern 1mu} {\gamma _{1,th}} + {h_{E2}} + Nv{h_2} - {h_{E1}}{\mkern 1mu} Nv } \right) \right. \right.\hfill \\
    \;\;\;\;\left. \left. \left. \left. \times {N_0} + P{h_2}{\mkern 1mu} {h_{E2}}\right) - {N_0}{\mkern 1mu} {\gamma _{1,th}}{\mkern 1mu} {h_2}{\mkern 1mu} {h_{E2}}\right)\right)^{\frac{1}{2}}\right) \hfill \\
    \;\;\;\;\;\;\times {\left( {1 + {\gamma _{1,th}}} \right)^{ - 1}}\left( - P{h_1}{\mkern 1mu} {h_2}{\mkern 1mu} \left( {1 + {\gamma _{1,th}}} \right)h_{E2}^2 - {h_2}{\mkern 1mu} {N_0}{\mkern 1mu} {h_{E2}} \right. \hfill \\
    \;\;\;\;\;\;\times Nv \left( {\left( {{h_1} - {h_{E1}}} \right){\gamma _{1,th}} + {h_1}} \right) + \left(\left( {1 + {\gamma _{1,th}}} \right){h_{E1}}{\mkern 1mu}\right. \hfill \\
    \;\;\;\;\;\;\times  {h_2}Nv{h_{E2}}{\mkern 1mu} \left( {\left( {Nv{N_0} + P{h_{E2}}} \right){h_1} - {N_0} } \right. \hfill \\
    \;\;\;\;\;\;\times \left.{h_{E2}}{\mkern 1mu} {\gamma _{1,th}}\right) \left({h_1} \left( \left( {{h_{E2}}{\mkern 1mu} {\gamma _{1,th}} + {h_{E2}} + Nv{h_2} - {h_{E1}}{\mkern 1mu} Nv} \right) \right. \right. \hfill \\
    \;\;\;\;\left. \left. \left. \left. \times {N_0} + P{h_2}{\mkern 1mu} {h_{E2}}\right) - {N_0}{\mkern 1mu} {\gamma _{1,th}}{\mkern 1mu} {h_2}{\mkern 1mu} {h_{E2}}\right)\right)^{\frac{1}{2}}\right)^{ - 1}. \hfill
\end{gathered}
&
\end{flalign*}

\end{appendices}

% that's all folks

\begin{thebibliography}{99}
\bibitem{r1}
``Cisco visual networking index: Global mobile data traffic forecast update, 20172022,'' Feb. 2019. [Online]. Available: https://www.cisco.com/c/en/us/solutions/collateral/service-provider/visual-networking-index-vni/white-paper-c11-738429.pdf

\bibitem{r2}
Y. Liu, Z. Qin, M. Elkashlan, Z. Ding, A. Nallanathan, and L. Hanzo, ``Nonorthogonal multiple access for 5G and beyond,'' \emph{Proc. of the IEEE}, vol. 105, no. 12, pp. 2347--2381, 2017.

\bibitem{r3}
A. Gatherer. (2018, June) What will 6G be? [Online]. Available: https://www.comsoc.org/publications/ctn/what-will-6G-be

\bibitem{r4}
W. Saad, M. Bennis, and M. Chen, ``A vision of 6g wireless systems: Applications, trends, technologies, and open research problems,'' \emph{IEEE Netw.}, vol. 34, no. 3, pp. 134--142, 2020.

\bibitem{r5}
M. D. Renzo, ``Keynote talk \#2: 6G wireless: Wireless networks empowered by reconfigurable intelligent surfaces,'' in \emph{Proc. IEEE Asia-Pacific Conf. Commun. (APCC)}, Ho Chi Minh City, Vietnam, pp. xxxiv--xxxv, 2019.

\bibitem{r6}
M. Di Renzo et al., ``Smart radio environments empowered by reconfigurable AI meta-surfaces: An idea whose time has come,'' \emph{EURASIP J. Wireless Commun. Netw.}, no. 129, pp. 1--20, 2019.

\bibitem{r7}
W. Yan, X. Yuan, Z. He, and X. Kuai, ``Passive beamforming and information transfer design for reconfigurable intelligent surfaces aided multiuser mimo systems,'' \emph{IEEE J. Sel. Areas Commun.}, vol. 38, no. 8, pp. 1793--1808, 2020.

\bibitem{r8}
Q. Wu and R. Zhang, ``Towards smart and reconfigurable environment: Intelligent reflecting surface aided wireless network,'' \emph{IEEE Commun. Mag.}, vol. 58, no. 1, pp. 106--112, 2020.

\bibitem{r9}
W. Tang, M. Z. Chen, X. Chen, J. Y. Dai, Y. Han, M. D. Renzo, Y. Zeng, S. Jin, Q. Cheng, and T. J. Cui, ``Wireless communications with reconfigurable intelligent surface: Path loss modeling and experimental measurement,'' 2019, \emph{arXiv: 1911.05326}.

\bibitem{r10}
T. Hou, Y. Liu, Z. Song, X. Sun, Y. Chen, and L. Hanzo, ``Reconfigurable intelligent surface aided noma networks,'' \emph{IEEE J. Sel. Areas Commun.}, vol. 38, no. 11, pp. 2575--2588, 2020.

\bibitem{r11}
C. Zhang, J. Ge, F. Gong, F. Jia, and N. Guo, ``Security-reliability tradeoff for untrusted and selfish relay-assisted D2D communications in heterogeneous cellular networks for IoT,'' \emph{IEEE Syst J.}, vol. 14, no. 2, pp. 2192--2201, 2020.

\bibitem{r12}
C. Zhang, J. Ge, J. Li, F. Gong, and H. Ding, ``Complexity-aware relay selection for 5G large-scale secure two-way relay systems,'' \emph{IEEE Trans. Veh. Technol.}, vol. 66, no. 6, pp. 5461--5465, 2017.

\bibitem{r13}
H. M. Furqan, J. M. Hamamreh, and H. Arslan, ``Physical layer security for noma: Requirements, merits, challenges, and recommendations,'' 2019, \emph{arXiv: 1905.05064}.

\bibitem{r14}
E. Basar, M. Di Renzo, J. De Rosny, M. Debbah, M. Alouini, and R. Zhang, ``Wireless communications through reconfigurable intelligent surfaces,'' \emph{IEEE Access}, vol. 7, pp. 116753--116773, 2019.

\bibitem{r15}
E. Bjornson, O. Ozdogan and  E. G. Larsson, ``Intelligent reflecting surface vs. decode-and-forward: How large surfaces are needed to beat relaying?'' 2019, \emph{arXiv: 1906.03949}.

\bibitem{r16}
C. Huang, A. Zappone, G. C. Alexandropoulos, M. Debbah, and C. Yuen, ``Reconfigurable intelligent surfaces for energy efficiency in wireless communication,'' \emph{IEEE Trans. Wireless Commun.}, vol. 18, no. 8, pp. 4157--4170, 2019.

\bibitem{r17}
Q. Wu and R. Zhang, ``Intelligent reflecting surface enhanced wireless network: Joint active and passive beamforming design,'' in \emph{Proc. IEEE Glob. Commun. Conf. (GLOBECOM)}, Abu Dhabi, United Arab Emirates, pp. 1--6, 2018.

\bibitem{r18}
Q. Wu and R. Zhang, ``Intelligent reflecting surface enhanced wireless network via joint active and passive beamforming,'' \emph{IEEE Trans. Wireless Commun.}, vol. 18, no. 11, pp. 5394--5409, 2019.

\bibitem{r32}
G. Zhou, C. Pan, H. Ren, K. Wang, M. D. Renzo and A. Nallanathan, ``Robust Beamforming Design for Intelligent Reflecting Surface Aided MISO Communication Systems,'' \emph{IEEE Wireless Commun. Lett.}, vol. 9, no. 10, pp. 1658--1662, 2020.

\bibitem{r33}
G. Zhou, C. Pan, H. Ren, K. Wang and A. Nallanathan, ``A Framework of Robust Transmission Design for IRS-Aided MISO Communications With Imperfect Cascaded Channels,'' \emph{IEEE Trans. Signal Process.}, vol. 68, pp. 5092--5106, 2020.

\bibitem{r34}
S. Hong, C. Pan, H. Ren, K. Wang, K. K. Chai and A. Nallanathan, ``Robust Transmission Design for Intelligent Reflecting Surface Aided Secure Communication Systems with Imperfect Cascaded CSI,'' 2020, \emph{arXiv: 2004.11580}.

\bibitem{r35}
C. Pan et al., ``Intelligent Reflecting Surface Aided MIMO Broadcasting for Simultaneous Wireless Information and Power Transfer,'' \emph{IEEE J. Sel. Areas Commun.}, vol. 38, no. 8, pp. 1719--1734, 2020.

\bibitem{r36}
C. Pan et al., ``Multicell MIMO Communications Relying on Intelligent Reflecting Surfaces,'' \emph{IEEE Trans. Wireless Commun.}, vol. 19, no. 8, pp. 5218--5233, 2020.

\bibitem{r19}
Z. Ding and H. Vincent Poor, ``A simple design of IRS-NOMA transmission,'' \emph{IEEE Commun. Lett.}, vol. 24, no. 5, pp. 1119--1123, 2020.

\bibitem{r20}
X. Mu, Y. Liu, L. Guo, J. Lin, and N. Al-Dhahir, ``Exploiting intelligent reflecting surfaces in NOMA networks: Joint beamforming optimization,'' \emph{IEEE Trans. Wireless Commun.}, vol. 19, no. 10, pp. 6884--6898, 2020.

\bibitem{r21}
M. Fu, Y. Zhou, and Y. Shi, ``Intelligent reflecting surface for downlink non-orthogonal multiple access networks,'' in \emph{2019 IEEE Globecom Workshops (GC Wkshps)}, pp. 1--6, 2019.

\bibitem{r22}
V. C. Thirumavalavan and T. S. Jayaraman, ``Ber analysis of reconfigurable intelligent surface assisted downlink power domain NOMA system,'' in \emph{Proc. Int. Conf. COMmun. Syst. NETworkS (COMSNETS)}, Bengaluru, India, pp. 519--522, 2020.

\bibitem{r23}
G. Yang, X. Xu, and Y. Liang, ``Intelligent reflecting surface assisted non-orthogonal multiple access,'' in \emph{Proc. IEEE Wireless Commun. Networking Conf. (WCNC)}, Seoul, Korea (South), pp. 1--6, 2020.

\bibitem{r37}
M. Zeng, X. Li, G. Li, W. Hao and O. A. Dobre, ``Sum Rate Maximization for IRS-assisted Uplink NOMA,'' \emph{IEEE Commun. Lett.}, vol. 25, no. 1, pp. 234--238, 2021.

\bibitem{r38}
S. Jiao, F. Fang, X. Zhou and H. Zhang, ``Joint Beamforming and Phase Shift Design in Downlink UAV Networks with IRS-Assisted NOMA,'' \emph{J. Commun. Inf. Netw.}, vol. 5, no. 2, pp. 138--149, 2020.

\bibitem{r24}
Z. Chu, W. Hao, P. Xiao, and J. Shi, ``Intelligent reflecting surface aided multi-antenna secure transmission,'' \emph{IEEE Wireless Commun. Lett.}, vol. 9, no. 1, pp. 108--112, 2020.

\bibitem{r25}
M. Cui, G. Zhang, and R. Zhang, ``Secure wireless communication via intelligent reflecting surface,'' \emph{IEEE Wireless Commun. Lett.}, vol. 8, no. 5, pp. 1410--1414, 2019.

\bibitem{r26}
L. Yang, Y. Jinxia, W. Xie, M. Hasna, T. Tsiftsis, and M. Di Renzo, ``Secrecy performance analysis of RIS-aided wireless communication systems,'' \emph{IEEE Trans. Veh. Technol.}, vol. 69, no. 10, pp. 12296--12300, 2020.

\bibitem{r27}
X. Yu, D. Xu, and R. Schober, ``Enabling secure wireless communications via intelligent reflecting surfaces,'' in \emph{Proc. IEEE Glob. Commun. Conf. (GLOBECOM)}, Waikoloa, HI, USA, pp. 1--6, 2019.

\bibitem{r28}
J. Chen, Y. Liang, Y. Pei, and H. Guo, ``Intelligent reflecting surface: A programmable wireless environment for physical layer security,'' \emph{IEEE Access}, vol. 7, pp. 82599--82612, 2019.

\bibitem{r29}
X. Guan, Q. Wu, and R. Zhang, ``Intelligent reflecting surface assisted secrecy communication: Is artificial noise helpful or not?,'' \emph{IEEE Wireless Commun. Lett.}, vol. 9, no. 6, pp. 778--782, 2020.

\bibitem{r39}
Q. Wang, F. Zhou, R. Q. Hu and Y. Qian, ``Energy-Efficient Beamforming and Cooperative Jamming in IRS-Assisted MISO Networks,'' in \emph{Proc. IEEE Int Conf Commun (ICC)}, Dublin, Ireland, pp. 1--7, 2020.

\bibitem{r30}
S. Hong, C. Pan, H. Ren, K. Wang and A. Nallanathan, ``Artificial-Noise-Aided Secure MIMO Wireless Communications via Intelligent Reflecting Surface,'' \emph{IEEE Trans. Commun.}, vol. 68, no. 12, pp. 7851--7866, 2020.

\bibitem{r31}
J. Hu, F. Shu and J. Li, ``Robust Synthesis Method for Secure Directional Modulation With Imperfect Direction Angle,'' \emph{IEEE Commun. Lett.}, vol. 20, no. 6, pp. 1084--1087, 2016.
{
\bibitem{r40}
Q. H. Spencer, A. L. Swindlehurst and M. Haardt, ``Zero-forcing methods for downlink spatial multiplexing in multiuser MIMO channels,'' \emph{IEEE Trans. Signal Process.}, vol. 52, no. 2, pp. 461--471, 2004.

\bibitem{r41}
Lai-U Choi and R. D. Murch, ``A transmit preprocessing technique for multiuser MIMO systems using a decomposition approach,'' \emph{IEEE Trans. Wirel. Commun.}, vol. 3, no. 1, pp. 20--24, 2004.

\bibitem{r42}
J. Li, J. Ge, C. Zhang, J. Shi, Y. Rui and M. Guizani, ``Impact of Channel Estimation Error on Bidirectional MABC-AF Relaying With Asymmetric Traffic Requirements,'' \emph{IEEE Trans. Veh. Technol.}, vol. 62, no. 4, pp. 1755--1769, 2013.
}

\end{thebibliography}
\end{document}